\documentclass[12pt,preprint]{aastex}

\shorttitle{Intrinsically red sources in the Galactic Plane}
\shortauthors{Robitaille et al.}

\usepackage{xspace}

\newcommand{\xagb}{xAGB\xspace}
\newcommand{\sagb}{sAGB\xspace}

\newcommand{\microns}{\,$\mu$m\xspace}
\newcommand{\mJy}{\,mJy\xspace}
\newcommand{\Jy}{\,Jy\xspace}
\newcommand{\MJysr}{\,MJy/sr\xspace}
\newcommand{\sqdeg}{\,deg$^2$\xspace}
\newcommand{\degrees}{$^\circ$\xspace}
\renewcommand{\arcsec}{$^{\prime\prime}$\xspace}
\renewcommand{\arcmin}{$^{\prime}$\xspace}
\newcommand{\kpc}{\,kpc\xspace}
\newcommand{\pc}{\,pc\xspace}
\newcommand{\seconds}{\,s\xspace}
\renewcommand{\mag}{\,mag\xspace}
\newcommand{\percent}{\%\xspace}
\newcommand{\ks}{K$_{\rm s}$\xspace}
\newcommand{\peryr}{\,yr$^{-1}$\xspace}
\newcommand{\au}{\,AU\xspace}
\newcommand{\lsun}{$L_{\odot}$\xspace}

\begin{document}

%%%%%%%%%%%%%%%%% Title Section %%%%%%%%%%%%%%%%%

\title{Intrinsically Red Sources observed by \textit{Spitzer} in the Galactic Mid-Plane}

\author{Thomas P. Robitaille\altaffilmark{1}, Marilyn R. Meade\altaffilmark{2}, Brian L. Babler\altaffilmark{2}, Barbara A. Whitney\altaffilmark{3}, Katharine G. Johnston\altaffilmark{1}, R\'emy Indebetouw\altaffilmark{4}, Martin Cohen\altaffilmark{5}, Matthew S. Povich\altaffilmark{2}, Marta Sewilo\altaffilmark{6}, Robert A. Benjamin\altaffilmark{2}, Edward Churchwell\altaffilmark{2}}

\email{tr9@st-andrews.ac.uk}

\altaffiltext{1}{SUPA, School of Physics and Astronomy, University of St Andrews, North 
Haugh, KY16 9SS, St Andrews, United Kingdom}
\altaffiltext{2}{Department of Astronomy, 475 North Charter St., University of Wisconsin, Madison, WI 53706}
\altaffiltext{3}{Space Science Institute, 4750 Walnut St. Suite 205, Boulder, CO 80301, USA}
\altaffiltext{4}{University of Virginia, Astronomy Dept., P.O. Box 3818, Charlottesville, VA, 22903-0818}
\altaffiltext{5}{Radio Astronomy Laboratory, 601 Campbell Hall, University of California at Berkeley, Berkeley, CA 94720}
\altaffiltext{6}{Space Telescope Science Institute, 3700 San Martin Way, Baltimore, MD 21218}

%%%%%%%%%%%%%%%%% Abstract %%%%%%%%%%%%%%%%%

\begin{abstract}

We present a highly reliable flux-limited census of 18,949 point sources in the Galactic mid-plane that have intrinsically red mid-infrared colors. These sources were selected from the \textit{Spitzer Space Telescope} GLIMPSE~I and II surveys of 274\sqdeg of the Galactic mid-plane, and consist mostly of high- and intermediate-mass young stellar objects (YSOs) and asymptotic giant branch (AGB) stars. The selection criteria were carefully chosen to minimize the effects of position-dependent sensitivity, saturation, and confusion. The distribution of sources on the sky and their location in IRAC and MIPS 24\microns color-magnitude and color-color space are presented. Using this large sample, we find that YSOs and AGB stars can be mostly separated by simple color-magnitude selection criteria into approximately $50-70$\percent of YSOs and $30-50$\percent of AGB stars. Planetary nebulae and background galaxies together represent at most $2-3$\percent of all the red sources. 1,004 red sources in the GLIMPSE~II region, mostly AGB stars with high mass-loss rates, show significant ($\ge$0.3\mag) variability at 4.5 and/or 8.0\microns. With over 11,000 likely YSOs and over 7,000 likely AGB stars, this is to date the largest uniform census of AGB stars and high- and intermediate mass YSOs in the Milky-Way Galaxy.

\end{abstract}


\maketitle

\section{Introduction}

The \textit{Spitzer Space Telescope} has recently completed a number of surveys of the Galactic mid-plane using the InfraRed Array Camera (IRAC; \citealt{Fazio:04:10}) at 3.6, 4.5, 5.8, and 8.0\microns, and the Multiband Imaging Photometer for \textit{Spitzer} (MIPS; \citealt{Rieke:04:25}) at 24 and 70\microns. In the context of star formation, the Galactic Legacy Infrared Mid-Plane Survey Extraordinaire \citep[GLIMPSE;][]{Benjamin:03:953} surveys - which to date includes GLIMPSE~I, GLIMPSE~II, and GLIMPSE 3D - and the MIPS Galaxy (MIPSGAL) surveys - which include the MIPSGAL I and II surveys - have so far been used for studies of individual star formation regions \citep[e.g.][]{Whitney:04:315,Indebetouw:07:321,Mercer:07:242,Shepherd:07:464,Povich:08}. However, the full potential of these surveys is that they provide a uniform view of the Galactic mid-plane - not only do they cover large and well-studied star formation regions, but they also show the distributed star formation between these regions. Therefore, in addition to focusing on specific regions, a whole continuum of star formation environments can now be studied. When seen in this light, these observations have the potential to revolutionize our view of Galactic star formation.

These surveys are not the first of their kind at mid- and far-infrared wavelengths. Previous major surveys covering the Galactic plane include the InfraRed Astronomical Satellite (IRAS; \citealt{Neugebauer:84:L1}) all-sky survey in 1983 at 12, 25, 60, and 100\microns, the Infrared Space Observatory (ISO) Galactic plane survey \citep{Omont:03:975} at 7 and 15\microns, and the Midcourse Space Experiment (MSX) survey of the Galactic plane \citep{Price:01:2819} at 8.28, 12.13, 14.65, and 21.3\microns. However, the combination of coverage, sensitivity and resolution of the \textit{Spitzer} observations is unprecedented: the full-width half-maximum (FWHM) of the point spread function (PSF) is 2\arcsec at 8\microns, and 6\arcsec at 24\microns, compared to the detector resolution of 18.3\arcsec for MSX 8.28 and 21.3\microns and a FWHM of approximately 3-5\arcmin for IRAS 12 and 25\microns. The point source sensitivity at 8\microns is 100 and 1,000 times better than MSX 8.28\microns and IRAS 12\microns respectively, and the sensitivity at 24\microns is also approximately 100 and 1,000 times more sensitive than MSX 21.3\microns and IRAS 25\microns respectively. At 7\microns, the sensitivity and resolution of the ISOGAL observations (9\mJy and 6\arcsec respectively) approach those of the \textit{Spitzer} GLIMPSE survey, but the coverage of the ISOGAL survey is only 6\percent of that of the GLIMPSE and MIPSGAL surveys (16\sqdeg for ISOGAL versus 274\sqdeg for GLIMPSE).

These three previous surveys have been used to search for young stellar objects (YSOs), which are brighter and redder at infrared wavelengths than field stars due to thermal emission from circumstellar dust. \citet{Wood:89:265} selected 1,717 candidate embedded massive stars with UCHII regions from the IRAS data, 1,646 of which are associated with the Galactic plane; \citet{Felli:02:20971} used the ISOGAL survey in conjunction with radio observations to compile a list of 715 YSO candidates; and the MSX survey was used to compile a Galaxy-wide sample of candidate massive YSOs which were followed up to eliminate contaminants via the Red MSX Source (RMS) survey \citep{Hoare:04:156}. However, the increased sensitivity of the recent \textit{Spitzer} observations of the Galactic mid-plane will allow many intermediate mass YSOs and more distant massive YSOs to be seen.

In this and future papers, our aim is to construct a photometrically reliable catalog of sources that is likely to contain many YSOs, and to use it to study the distribution of star formation regions in the Galaxy, the environments in which stars form, and to estimate the present rate of star formation in the Galaxy. The current paper describes the initial compilation of a red source catalog which is photometrically very reliable, and is affected as little as possible by any biases due for example to position-dependent sensitivity or saturation limits.
In addition to YSOs, a number of asymptotic giant branch (AGB) stars, which are also red at mid-infrared wavelengths due to the dust that surrounds them, are present in the red source catalog presented here. Because interstellar extinction is low at mid-infrared wavelengths, and because of the selection criteria used for this catalog, these are amongst the reddest and most distant AGB stars in the Galaxy.

In \S\ref{sec:observations}, we describe the GLIMPSE observations that are used to compile the red source catalog, the various issues that affect the completeness of the GLIMPSE Catalogs, and the complementary surveys that are used to construct SEDs from 1.25 to 24\microns. In \S\ref{sec:selection}, we describe the selection of the red sources from the GLIMPSE Catalog, including the procedure used to increase the reliability and improve the uniformity of the selection across the Galaxy. In \S\ref{sec:analysis} we show the angular distribution and colors of the red sources, we identify over 1,000 red variable sources, and we study the composition of the red source catalog. Finally, in \S\ref{sec:summary}, we summarize our findings.

\section{Observations}

\label{sec:observations}

\subsection{Description of the IRAC observations}

\label{sec:irac_description}

\begin{figure}[p]
\begin{center}
\includegraphics[width=6in]{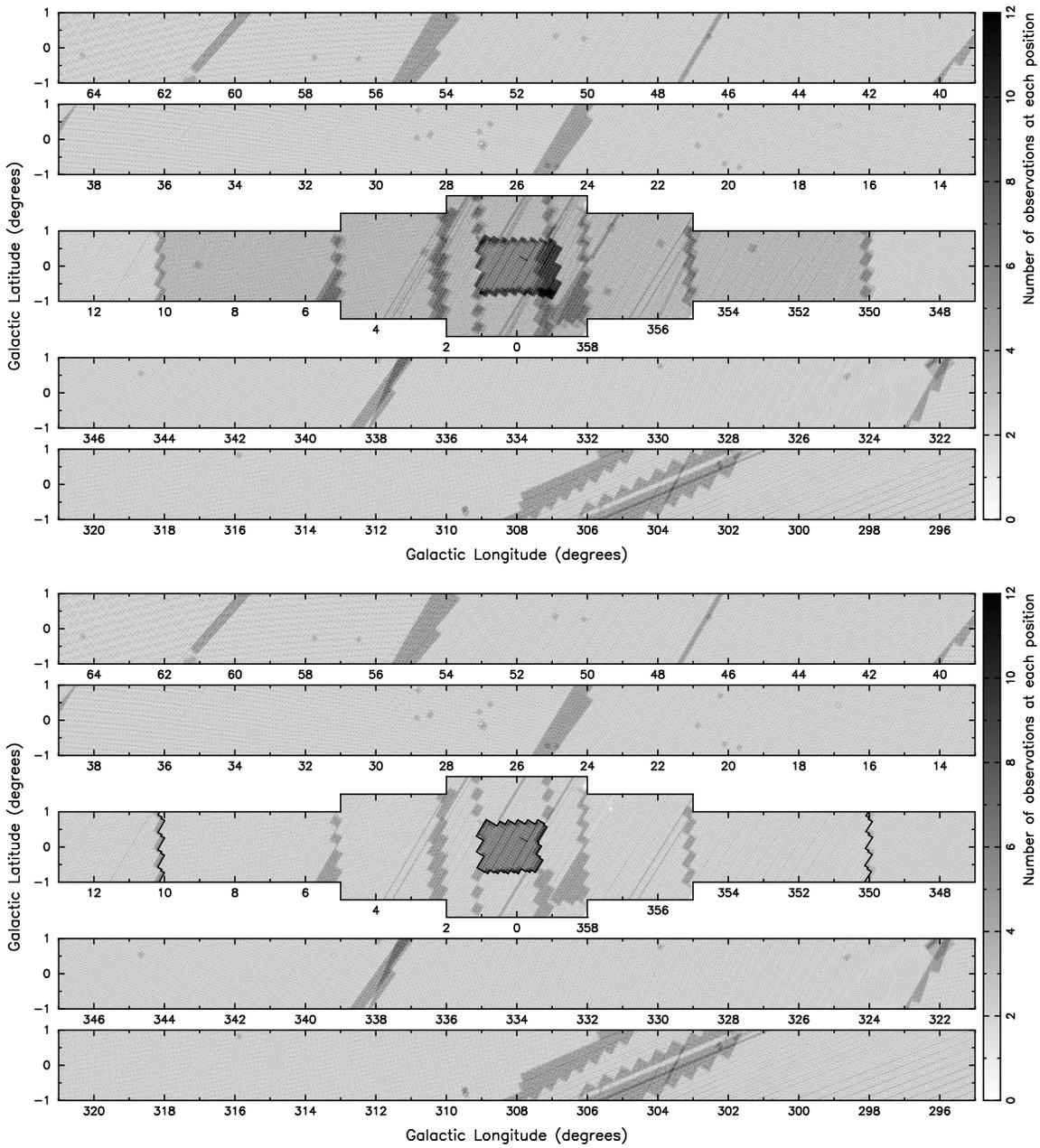}
\caption{\small Coverage of the GLIMPSE~I and II surveys, and the Galactic center data.  The top panel shows the coverage using all available GLIMPSE observations, while the bottom panel shows the coverage if only the first epoch data is used inside the GLIMPSE~II region. This region is enclosed by the thick lines at $\ell=10$\degrees and $\ell=350$\degrees, and excludes the Galactic center region, also enclosed by a thick line. The scale used is shown on the right in each case: darker shades of gray indicate areas that have been observed a larger number of times.\label{fig:irac_coverage}}
\end{center}
\end{figure}

In this paper, we make use of observations taken with the \textit{Spitzer} IRAC camera to select intrinsically red sources. The IRAC data used are from the GLIMPSE~I survey (PI: Churchwell; PIDs 146, 186, 187, 188, 189, 190, 191, 192, and 195), the GLIMPSE~II survey of the inner region of the Galactic plane (PI: Churchwell; PID 20201), and observations of the Galactic center (PI: Stolovy; PID 3677). GLIMPSE~I covers $10$\degrees $\le|\ell|\le65$\degrees and $|b|\le1$\degrees. GLIMPSE~II fills in the region for $|\ell|<10$\degrees, with $|b|\le1$\degrees for $|\ell|>5$\degrees , $|b|\le1.5$\degrees for 2\degrees$<|\ell|\le5$\degrees, and $|b|\le2$\degrees for $|\ell|\le2$\degrees. The total area surveyed is thus 274\sqdeg. In reality, due to the observing strategy, the surveys extend slightly beyond these limits; however throughout the remainder of this paper only sources inside
this `official' survey area are considered, as this makes calculations of surface
densities of sources and the analysis of clustering more straightforward. The coverage of all the IRAC observations is shown in the top panel of Figure~\ref{fig:irac_coverage}.

All IRAC observations consist of frames with 1.2\seconds effective exposure times. In principle, GLIMPSE~I has one epoch of observations, with two exposures at each position, GLIMPSE~II has two epochs of observations (separated by six months), with three exposures at each position (two during the first epoch, and one during the second), and the Galactic center data have a single epoch of observations, with five exposures at each position. In practice, a given position can be covered more often than this because of re-observed missing or bad frames, and overlap between the individual frames, between the observing campaigns, and between the different surveys. This can be seen in Figure~\ref{fig:irac_coverage}.

The GLIMPSE~I and II v2.0 all-epoch enhanced data products consist of highly reliable Point Source Catalogs, more complete Point Source Archives, and mosaic images with both 0.6\arcsec and 1.2\arcsec pixel resolutions. This processing of the all-epoch GLIMPSE~II data includes all GLIMPSE~I basic calibrated data (BCD) frames for $|\ell|<11$\degrees, and includes all of the Galactic center data, and therefore supersedes the GLIMPSE~I v2.0 data products for $9^\circ<|\ell|<11$\degrees. Since the GLIMPSE~II survey is two epoch, single-epoch data products are also available for the whole GLIMPSE~II area. For example, the first-epoch enhanced data products include only GLIMPSE~II BCD frames from the first epoch (and exclude GLIMPSE~II second-epoch, GLIMPSE~I, or Galactic center data).

For this work, the Point Source Catalogs were used, as they have a higher reliability than the Point Source Archives. The Catalogs are high-quality subsets of the Archives: for example, one of the main differences between the Catalogs and the Archives is that a source can be included in the Archives if it is detected only in one IRAC band, whereas it is required to be detected in at least two neighboring bands in order to be included in the Catalogs. In addition, some fluxes present for a given source in the Archives may be nulled in the Catalogs, for example if the fluxes approach the saturation levels. More details on the GLIMPSE data products, such as the selection criteria for the Archives and Catalogs, are provided in the GLIMPSE~I and II Science Data Products Document\footnote{\url{http://irsa.ipac.caltech.edu/data/SPITZER/GLIMPSE/doc/glimpse1\_dataprod\_v2.0.pdf\label{foot:products}}}$^{,}$\footnote{\url{http://irsa.ipac.caltech.edu/data/SPITZER/GLIMPSE/doc/glimpse2\_dataprod\_v2.0.pdf\label{foot:products2}}} and the GLIMPSE Quality Assurance Document\footnote{\url{http://irsa.ipac.caltech.edu/data/SPITZER/GLIMPSE/doc/glimpse\_quality\_assurance\_v1.0.pdf\label{foot:quality}}}. As will be described in \S\ref{sec:comp_variability}, highly variable sources will be missing from the all-epoch Catalogs in the GLIMPSE~II survey area in particular, so the first-epoch Catalogs were used instead of the all-epoch Catalogs to select sources from the GLIMPSE~II region.

\subsection{Completeness of the IRAC observations}

\label{sec:completeness}

In this section, we review the effects of position dependent saturation and sensitivity, variability, and confusion on the completeness of the GLIMPSE~I and II surveys. As will be discussed in \S\ref{sec:qualitative_selection}, only IRAC 4.5 and 8.0\microns fluxes were used for the source selection. Therefore, the following discussions refer mostly to these two bands.

\subsubsection{Saturation}

Since all IRAC observations used the same effective exposure time of 1.2\seconds, the \textit{pixel} saturation level is independent of position across the whole survey. However, the maximum flux a point source can have without being saturated will depend on the background emission. For instance, if the background is very bright, and close to saturation, only faint point sources will avoid saturation. Since the level of diffuse emission is a strong function of position in the survey area, the \textit{point source} flux saturation level will be position-dependent. Using the equation for point source saturation from the \textit{Spitzer Observer's Manual}\footnote{\url{http://ssc.spitzer.caltech.edu/documents/som/}}, the equation for the saturation flux at 4.5\microns is
\begin{equation}
\frac{F_{\rm sat}}{\rm mJy} = \frac{1}{11.09}\left(3621 - \frac{B}{\rm MJy/sr}\right),
\end{equation}
and for 8.0\microns, the equation is
\begin{equation}
\frac{F_{\rm sat}}{\rm mJy} = \frac{1}{4.79}\left(5878 - \frac{B}{\rm MJy/sr}\right),
\end{equation}
where $F_{\rm sat}$ is the point source saturation flux and $B$ is the background level. In the ideal case of a zero background level, the point source saturation levels in these two bands are therefore 327 and 1228\mJy respectively. These are conservative `worst-case' values, which assume for example that the sources are perfectly aligned with the center of pixels. In reality, it was found in the GLIMPSE processing that fluxes up to 450 and 1,590\mJy could be extracted reliably at 4.5 and 8.0\microns respectively, i.e. 30-40\percent higher than the \textit{Spitzer} Science Center (SSC) 'worst case' values. However, the dependence of the point source saturation limit is essentially correct: as the background level increases, the point source saturation flux decreases.

At 4.5\microns, the diffuse emission (excluding the zodiacal light) almost never exceeds 25\MJysr, which means that in practice, the point source saturation flux will change by less than 1\percent as a function of position. The only notable exceptions where the diffuse emission is brighter than this (and exceeds 100\MJysr) are the Galactic center and the Omega Nebula (M17) star formation region.

At 8.0\microns, in regions where the diffuse emission is as high as 1,000\MJysr, all sources fainter than $\sim$1,000\mJy should still be detectable. In practice, such a high background value is very rare: the fraction of the survey for which the diffuse emission brightness is above 1,000\MJysr is approximately 0.015\percent (approximately 0.04\sqdeg). Furthermore, only very few sources (135) in the final red source catalog (0.7\percent) are brighter than 1,000\mJy at 8.0\microns. Therefore, the probability of a bright ($>$1,000\mJy) source being in a region of bright ($>$1,000\MJysr) diffuse emission is very low. This means that the dependence of the saturation limit on the diffuse emission brightness is likely to remain unnoticed for this work. The exception to this is the M17 star formation region, which is so bright at 8\microns that the diffuse emission saturates in places; thus the point source saturation flux in this region decreases substantially, and reaches zero where the diffuse emission saturates.

\subsubsection{Sensitivity}

\label{sec:sensitivity}

\begin{figure}[t]
\begin{center}
\includegraphics[width=6in]{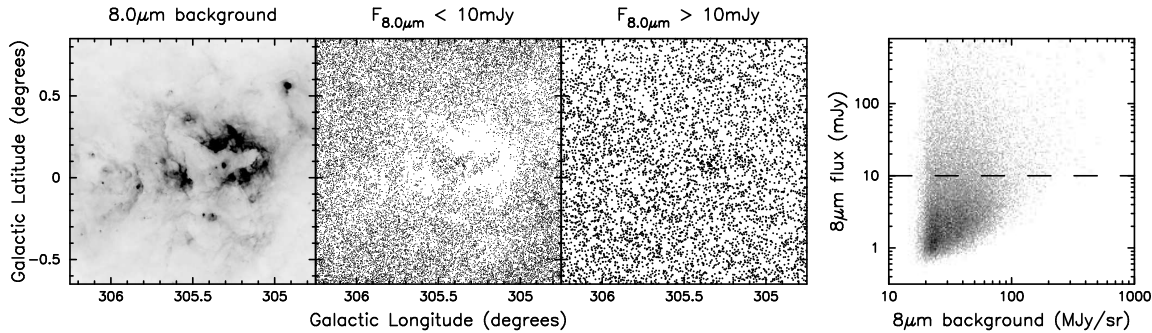}
\caption{The dependence of sensitivity on the diffuse emission brightness. From left to right: 8.0\microns diffuse emission intensity for a 1.5\degrees$\times$1.5\degrees field centered on ($\ell$,$b$) = (305.5\degrees,0.1\degrees) (where a darker shade of gray indicates brighter diffuse emission); point sources (shown as black points) extracted with 8.0\microns flux densities less than 10\mJy; point sources (shown as black points) extracted with 8.0\microns flux densities more than 10\mJy; flux density versus diffuse emission brightness for all the sources in this field (shown on a grayscale where darker shades of gray indicate more sources). The horizontal dashed line corresponds to $F_\nu=10$\mJy. Only point sources satisfying the brightness and quality selection criteria from equations (\ref{eq:brightness}) and (\ref{eq:quality}) are used in this figure.\label{fig:sens_complete}}
\end{center}
\end{figure}

Although different regions of the survey were observed a different number of times, the photometry for the GLIMPSE~I and II catalogs was always done on 1.2\seconds BCD frames, meaning that the number of observations at a given position should not in principle affect the sensitivity limit (this would be different if the source detection and photometry were both carried out on mosaics). In practice, the more a source is observed, the more likely it is to be detected enough times to satisfy the Catalog selection criteria, but this effect is not found to be dominant: for example, there is no jump in the number of sources at $\ell=10$\degrees or $\ell=350$\degrees between the all-epoch GLIMPSE~I and II Catalogs, despite the fact that the all-epoch GLIMPSE~II Catalogs use three observations at each position instead of two.

The main factor that determines the sensitivity limit, in particular at 8.0\microns, is the brightness of the diffuse emission. The brighter the background emission, the larger the Poisson noise, and therefore the poorer the sensitivity. This has a strong effect on the detectability of sources, mainly for sources fainter than 10\mJy at 8\microns. To illustrate this, the distribution of sources around the $\ell$=305\degrees region is shown in Figure~\ref{fig:sens_complete}, distinguishing between sources brighter and sources fainter than 10\mJy (only sources satisfying the data quality criteria outlined in \S\ref{sec:initial_selection} are shown). The right hand panel clearly shows the dependence of the faintest source detected as a function of background flux. To ensure that the angular distribution of sources in our red source catalog is not affected by the variations in the sky background, lower limits on the 4.5 and 8.0\microns fluxes of 0.5 and 10\mJy respectively ([4.5]=13.89 and [8.0]=9.52) will be imposed in \S\ref{sec:initial_selection}.

\subsubsection{Variability}

\label{sec:comp_variability}

The issue of variability between the various epochs of observations is important, as the entire survey area is constructed from BCD frames taken at different epochs, separated in some cases by over a year. The bandmerger in the GLIMPSE pipeline, a modified version of the SSC bandmerger\footnote{http://ssc.spitzer.caltech.edu/postbcd/bandmerge.html}, makes uses of the positions, fluxes, and flux uncertainties of detections within the same band, but in different BCD frames, in order to determine whether to combine the detections. The larger the positional offset between two detections, the more the fluxes have to agree for the detections to be combined, and vice-versa. Thus, two detections with differing fluxes are most likely to be matched if they are positionally coincident. In this case, to be combined, the fluxes can differ by up to 4.5$\sigma_{\rm total}$, where $\sigma_{\rm total}^2=\sigma_1^2+\sigma_2^2$, and $\sigma_1$ and $\sigma_2$ are the flux uncertainties of the two detections respectively. The threshold of 4.5$\sigma$ was determined empirically by running the GLIMPSE bandmerger with various combinations of nearby detections and a range of different fluxes. Therefore, detections with fluxes differing by more than 4.5$\sigma_{\rm total}$ are never combined regardless of their positional offset. Not combining two detections results in the detections being treated as two separate sources, which in turns decreases the chance of either of these sources making it to the final Archive or Catalog. No constraints are placed on the fluxes when deciding to consider detections in {\it different bands} as belonging to the same source (that decision is based on close positional coincidence only).

For the work presented here, sources are only selected from the Catalogs if the standard deviation of the fluxes from the different BCD frames is less than 15\percent of the mean flux, as will be described in \S\ref{sec:initial_selection}. Therefore, although it is not straightforward to know whether a given variable source will be included in the GLIMPSE Catalogs due to the merging of detections in the GLIMPSE pipeline, the requirement for the standard deviation to be less than 15\percent does remove any sources that are significantly variable from regions covered at several epochs in the all-epoch Catalogs. We note that no variable stars are removed in regions covered only at a single epoch (i.e. most of GLIMPSE~I and the Galactic center).

\begin{figure}[p]
\begin{center}
\includegraphics[width=6in]{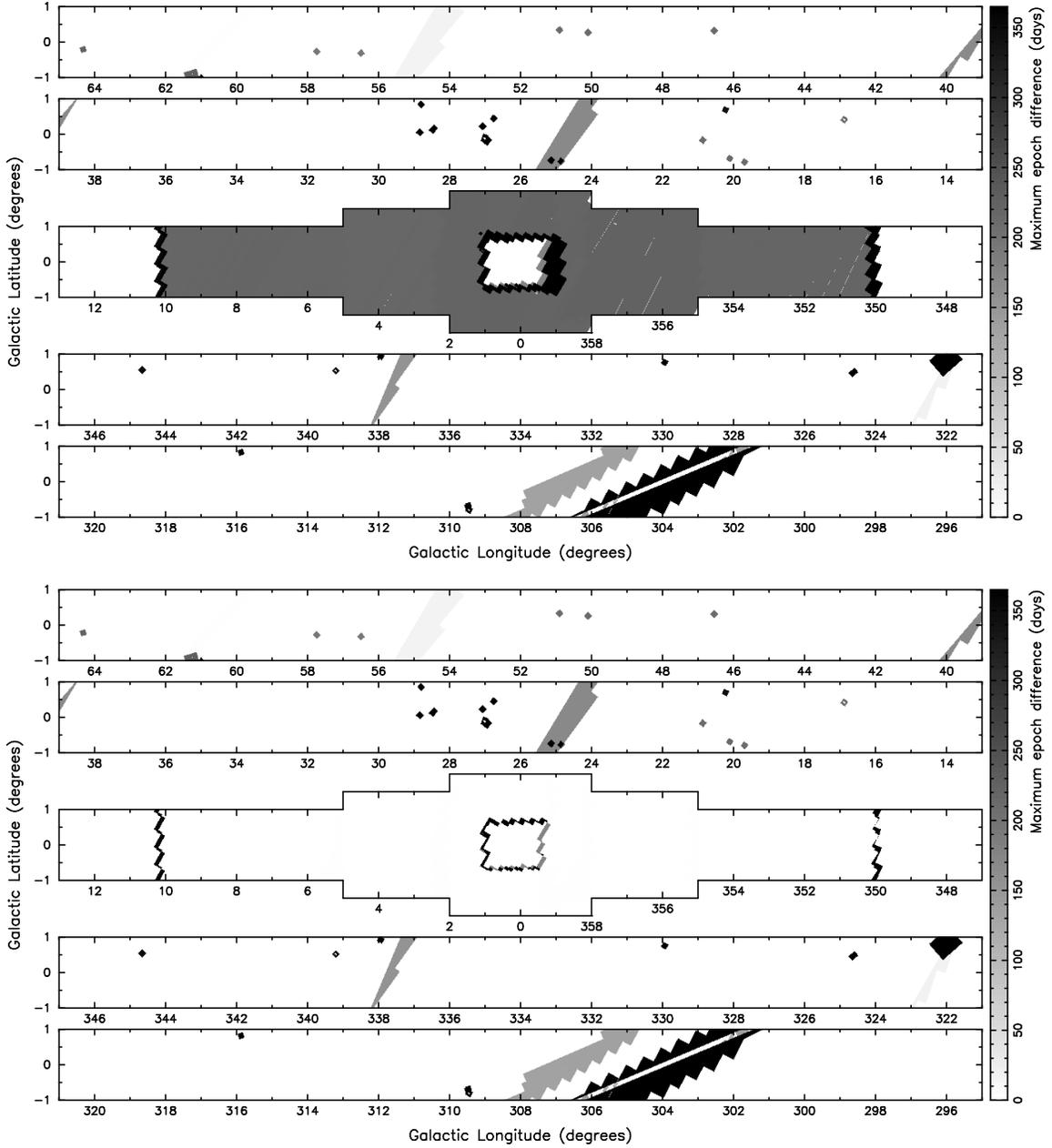}
\caption{\small Maximum epoch difference between BCD frames as a function of position. The top panel shows the epoch difference in the BCD frames used for the all-epoch data products, while the bottom panel uses exclusively the GLIMPSE~II first epoch BCD frames where available, and all BCD frames elsewhere. The scale used is shown on the right in each case: darker shades of gray correspond to larger epoch differences, with differences of a year and above shown in black. Therefore, white corresponds to areas covered effectively at a single epoch.\label{fig:var_complete}}
\end{center}
\end{figure}

In Figure~\ref{fig:var_complete}, the maximum difference in epochs between BCD frames is shown as a function of position. The top panel shows this using all BCD frames available, and is therefore a map of where variable sources are likely to be missing when using the all-epoch v2.0 Catalogs. The GLIMPSE~I area is mostly single epoch, as the two exposures at each position were taken 20 seconds apart. GLIMPSE~I was observed in segments of 15\degrees of longitude, separated by intervals of weeks to months, so the regions of overlap between these segments are effectively multi-epoch (the epoch difference is 15 to 20 days at  $\ell=55$\degrees and $322$\degrees, and is 130 to 170 days at $\ell=40$\degrees, $25$\degrees, $337$\degrees, and $307$\degrees). A small fraction of the survey was re-observed to fill in gaps in the coverage, resulting in small multi-epoch patches. The largest re-observed region is the region between $\ell$=302\degrees and 306\degrees, which was re-observed 611 days after the original survey. The Galactic center observations are single-epoch, as the entire set of observations was completed within 16 hours. Finally, the GLIMPSE~II area is two-epoch by design, with the two epochs separated by 215 to 225 days. Since the two GLIMPSE~II epochs were observed after GLIMPSE~I and after the Galactic center observations, the regions of overlap between GLIMPSE~II, GLIMPSE~I and the Galactic center data were observed at three different epochs, with a maximum epoch difference of  597 days at the overlap region between GLIMPSE~I and II (at $|\ell|=10$\degrees), and 396 days at the overlap region between GLIMPSE~II and the Galactic center observations (at $|\ell|=1$\degrees and $|b|<0.75$\degrees, and at $|b|=0.75$\degrees and $|\ell|<1$\degrees). The bottom panel of Figure~\ref{fig:var_complete} shows the same as the top panel with the exception that wherever first-epoch GLIMPSE~II BCD frames are available, only those are used; this region is outlined in the bottom panel of Figure~\ref{fig:irac_coverage}. This dramatically reduces the area in which variable sources will be missing.

To summarize, a large fraction of significantly variable stars is likely to be absent from the GLIMPSE all-epoch Catalogs in regions covered at several epochs (for epochs sufficiently far apart). These regions occur (a) within each individual survey, in regions of overlap between different observing campaigns, or where re-observations have been carried out, (b) in the entire GLIMPSE~II survey region, since this survey was two-epoch by design, and (c) in the regions of overlap between GLIMPSE~I and II, and between GLIMPSE~II and the Galactic center data. However, by using the GLIMPSE~II first-epoch Catalogs instead of the all-epoch Catalogs wherever possible, the number of excluded variables sources can be minimized.
 
\subsubsection{Confusion}

\label{sec:confusion}

As described in the GLIMPSE Quality Assurance document$^3$, photometry can become unreliable when two sources are separated by less than 2.4\arcsec. In this situation, `flux stealing' - splitting the flux incorrectly between two sources - becomes important. This is most likely to happen in dense clusters and in the GLIMPSE~II survey area as one approaches the Galactic center. All sources in the GLIMPSE Catalog are assigned a \textit{close source flag} that indicates whether another source is closer than 2.5\arcsec (flag set to 2), closer than 3.0\arcsec (flag set to 1), or whether there is no other source within 3.0\arcsec (flag set to 0). Since photometric accuracy is very important for this work, only sources that have no neighbor closer than 3.0\arcsec are kept (as described further in \S\ref{sec:initial_selection}). However, this means that the red source catalog will be incomplete in regions of very high stellar densities such as dense clusters or towards the Galactic center.

\subsection{Description of the complementary observations}

For this work, complementary observations were used to construct SEDs from 1.25 to 24\microns for the sources selected as intrinsically red. These complementary observations are:

\begin{itemize}

\item The 2MASS Point Source Catalog \citep{Skrutskie:06:1163}, which includes photometry in  JH\ks filters for the whole area covered by the GLIMPSE surveys. The 2MASS photometry is merged with the IRAC data in the GLIMPSE pipeline, and is listed in the GLIMPSE~I and GLIMPSE~II catalogs. The 10-$\sigma$ sensitivity limits are typically 15.8, 15.1, and 14.3\mag for J, H, and \ks respectively.

\item \textit{Spitzer} MIPS data from the MIPSGAL I survey of the GLIMPSE~I area, (PI: Carey; PID 20597), the MIPSGAL II survey of the GLIMPSE~II area (PI: Carey; PID 30594), and observations of the Galactic center (PI: Yusef-Zadeh; PID 20414). Since Point Source Catalogs are not available at the time of publication, the post-basic calibrated data (PBCD) mosaics were used to perform the photometry.

\item MSX band E (21.3\microns) data for sources saturated at MIPS 24\microns.

\end{itemize}

\section{Source selection and catalog compilation}

\label{sec:selection}

\subsection{Definition of an intrinsically red source}

\label{sec:qualitative_selection}

Due to the difficulty in separating YSOs, AGB stars, planetary nebulae (PNe), and other red sources at mid-infrared wavelengths, we decided to first create a catalog of all intrinsically red sources, and to leave the separation of the various populations until after compiling the catalog. The definition of an \textit{intrinsically red source} that we adopt is one that is intrinsically redder at IRAC wavelengths than field stars (such as main sequence or red giant stars), and would therefore still be red in the absence of interstellar extinction.

One option to select red sources would be to remove sources that could be well fit by reddened stellar photosphere models, allowing interstellar extinction to be a free parameter, and considering the remaining sources to be intrinsically red \citep[e.g.][]{Indebetouw:07:321,Poulton:08:1249}. However, while SED modeling can usually provide unique insights into the properties of objects by making the best use of multi-wavelength data, such a procedure would not be suitable for generating a red source catalog for several reasons. Firstly, the goodness of fit is quantified by a $\chi^2$ value, which not only depends on the fluxes of a source, but also the flux errors, meaning that the number of sources selected is very sensitive to the $\chi^2$ threshold and the choice of flux errors. In addition, the number of sources badly fit by reddened photospheres strongly depends on the specific stellar photosphere models used. For example, using \citet{Castelli:04} models resulted in approximately twice as many remaining red sources as the \citet{Brott:05:565} models for a same $\chi^2$ cutoff. Finally, such a selection is not easily reproducible. Undoubtedly, the models will improve in the future, and might result in yet a different number of red sources given the same selection criterion. Reproducing the selection criterion would require using the same version of the models as was used in this paper.

Instead, we decided to extract intrinsically red sources using a color selection. A multi-color selection criterion - combined with different sensitivity and saturation limits at different wavelengths - would make it difficult to understand selection biases. Therefore, a very simple single-color selection criterion was chosen, namely that the color between IRAC 4.5\microns and IRAC 8.0\microns be above a certain threshold for a source to be considered as red. While this selection may appear simplistic at first, it allows a much better understanding of selection biases, and is easily reproducible. A cutoff value of $[4.5]-[8.0]\ge1$ was used to select red sources, and this choice is justified in \S\ref{sec:red_selection}.

The choice of IRAC 4.5\microns (rather than 3.6\microns) as the lower wavelength was motivated by recent results which suggest that the interstellar extinction law is approximately flat between 4.5\microns and 8.0\microns \citep{Lutz:99:623,Indebetouw:05:931,Flaherty:07:1069}. Using the selective extinction values from \citeauthor{Indebetouw:05:931} and \citeauthor{Flaherty:07:1069}, we derive $E([4.5]-[8.0])/A_K=0.000\pm0.040$ and $E([4.5]-[8.0])/A_K=0.041\pm0.020$ respectively, where the uncertainties are the standard deviations of the values for the different lines of sight, and effectively represent the region to region variations. Therefore, a star with an intrinsic color of $[4.5]-[8.0]=0$ would need to be seen through an extinction of at least $A_{\rm K}=25$ to have $[4.5]-[8.0]\ge1$, that is $A_{\rm V}\approx190-220$ assuming $A_{V}/A_K\approx7.5-8.8$ \citep{Cardelli:89:245}. Even if the intrinsic color of a source was $[4.5]-[8.0]=0.5$ (for example due to an absorption line or band at 4.5\microns), this would still require $A_{\rm V}\approx100$ to obtain $[4.5]-[8.0]\ge1$. While such high extinctions can occur through dark clouds for example, the vast majority of GLIMPSE sources will not be subjected to such high interstellar extinctions.

\subsection{Initial selection criteria}

\label{sec:initial_selection}

As described in \S\ref{sec:sensitivity}, the variations in the diffuse emission throughout the survey translate into a different point source sensitivity as a function of position. Since the aim was to produce a complete catalog within color and magnitude selection criteria, only sources that had 4.5\microns fluxes equal to or larger than 0.5\mJy, and 8.0\microns fluxes equal to or larger than 10\mJy were selected.

The issue of contamination by bad photometry and erroneous detections required the most attention when compiling the catalog. The source detection reliability for the GLIMPSE Catalogs is required to be at least 99.5\percent, and the overall accuracy of the photometry is also very high. However, applying a color selection can lead to a bias towards selecting sources with erroneous fluxes: for example, if a given source has either its 4.5 or the 8.0\microns flux erroneously estimated, this source will have non-stellar colors; in particular, sources with over-estimated 8.0\microns fluxes or underestimated 4.5\microns fluxes have a higher likelihood of being selected by a criterion such as $[4.5]-[8.0]\ge1$ than sources with accurate photometry. Therefore, it is important for this work to ensure that the photometry is as accurate as possible.

There are a number of reasons why fluxes might be wrongly estimated for a given source, one of which - `flux stealing' by close neighbors - was already mentioned in \S\ref{sec:confusion}. Since close neighbors can lead to uncertain photometry, only sources with a close source flag set to zero were selected (no Archive source within 3\arcsec).

Another possible source of unreliable photometry is low signal-to-noise detections. One common cause of over-estimation of fluxes occurs when a spurious local peak in the background emission (due to noise or cosmic rays) is mistaken for a point source, or when a source that should not have been detected becomes bright enough to be detected due to Poisson fluctuations in the number of photons from that source (Malmquist bias). In addition, genuine sources with low signal-to-noise may also have their fluxes wrongly estimated. Assuming that the photometric errors are correctly computed, the reliability of the Catalogs can be increased by applying signal-to-noise cuts. For the purposes of the work, only sources with fractional flux errors below 15\percent were selected. 

For a small fraction of sources, photometric uncertainties may be under-estimated, and unreliable photometry may remain. Spurious detections are not likely to happen twice at the same position in two different observations, and can therefore be identified by looking for sources detected only in one BCD frame, in regions covered by multiple BCD frames. In order to eliminate such sources, only sources detected at least twice at 4.5 and 8.0\microns were selected. For genuine faint sources with low signal-to-noise, one can use the standard deviation of the fluxes from the multiple detections to check whether the flux is reliable. For this reason, the standard deviation of detections was required to be less than 15\percent of the fluxes.

\begin{figure}[t]
\begin{center}
\includegraphics[width=6in]{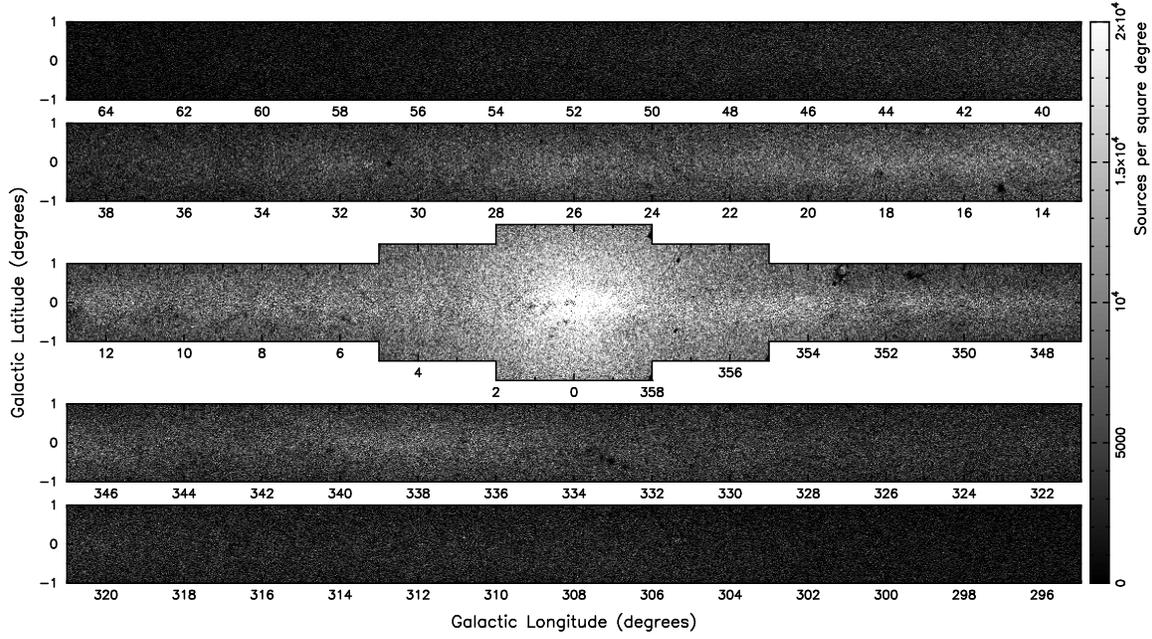}
\caption{Angular distribution of the GLIMPSE~I and II Catalog sources satisfying equations (\ref{eq:brightness}) and (\ref{eq:quality}). The scale used is shown on the right: lighter shades of grey correspond to a higher stellar density.\label{fig:spatial_all}}
\end{center}
\end{figure}

As discussed in \S\ref{sec:comp_variability}, in order to eliminate as few variable stars as possible, the GLIMPSE~II first-epoch Catalogs were used for the GLIMPSE~II first epoch survey area, and the GLIMPSE~I and II all-epoch Catalogs were used for the remaining area, i.e. most of the GLIMPSE~I and Galactic center regions. The initial selection criterion applied to these Catalogs can be summarized into two criteria: \textit{brightness} criteria, and \textit{quality} criteria. The brightness criteria are
\begin{equation}
\label{eq:brightness}
\left\{\begin{array}{rccclcrcccl}
0.5\,{\rm mJy} &\le& F_{4.5\,\mu{\rm m}} &\le& 450\,{\rm mJy}, & {\rm i.e.} & 13.89 &\ge& [4.5] &\ge& 6.50\\
10\,{\rm mJy} &\le& F_{8.0\,\mu{\rm m}} &\le& 1,590\,{\rm mJy}, & {\rm i.e.} & 9.52 &\ge& [8.0] &\ge& 4.01
\end{array}\right. .
\end{equation}
and the quality criteria are
\begin{equation}
\label{eq:quality}
\left\{\begin{array}{ll}
{\rm csf}=0&\\
dF_i/F_i \le 15\percent &i=2,4\\
M_i \ge 2 &i=2,4\\
F_i\_{\rm rms}/F_i \le 15\percent &i=2,4
\end{array}\right.
\end{equation}
using the notation from the GLIMPSE Science Data Products Documents$^{1,2}$: csf is the close source flag, $F_i$ and $dF_i$ are the fluxes and 1$\sigma$ errors, $M_i$ is the number of detections, $F_i\_$rms is the RMS or standard deviation of individual detections from $F_i$, and i is the IRAC band number, where $i=2$ corresponds to 4.5\microns, and $i=4$ corresponds to 8.0\microns.

The number of sources selected from the GLIMPSE~I and II all-epoch Catalogs (excluding the GLIMPSE~II first epoch area) and from the GLIMPSE~II first-epoch Catalogs after each requirement are listed in Table~\ref{tab:glimpse}. The distribution of the sources after applying the brightness and quality selection criteria is shown in Figure~\ref{fig:spatial_all}. A few regions show a decrease in the surface density of sources due to the diffuse emission (e.g. M17 at $\ell$=15\degrees, and the regions at $\ell$=333, 351, and 353\degrees) despite removing all 8.0\microns fluxes below 10\mJy, but the distribution of sources seems otherwise unaffected by diffuse emission.

\begin{deluxetable}{lcccc}
\tabletypesize{\footnotesize}
\tablewidth{0pt}
\tablecaption{Numbers of sources in the GLIMPSE Catalogs after selection criteria.\label{tab:glimpse}}
\tablehead{Selection criteria & GLIMPSE~I + GC\tablenotemark{a} & GLIMPSE~II & Total\\
 & all-epoch & first epoch}
\startdata
Eq. (\ref{eq:brightness}) & 908,748 & 591,772 & 1,500,520 \\
Eqs. (\ref{eq:brightness}) and (\ref{eq:quality}) & 828,795 & 526,587 & 1,355,382 \\
Eqs. (\ref{eq:brightness}), (\ref{eq:quality}) and (\ref{eq:color}) & 17,104 & 4,995 & 22,099\\
\enddata
\tablenotetext{a}{Excluding sources that lie in the GLIMPSE~II first epoch survey area}
\tablecomments{Only sources inside the `official' survey area are included, as described in \S\ref{sec:irac_description}}
\end{deluxetable}

\subsection{Selection of red sources}

\label{sec:red_selection}

\begin{figure}[t]
\begin{center}
\includegraphics[width=6.0in]{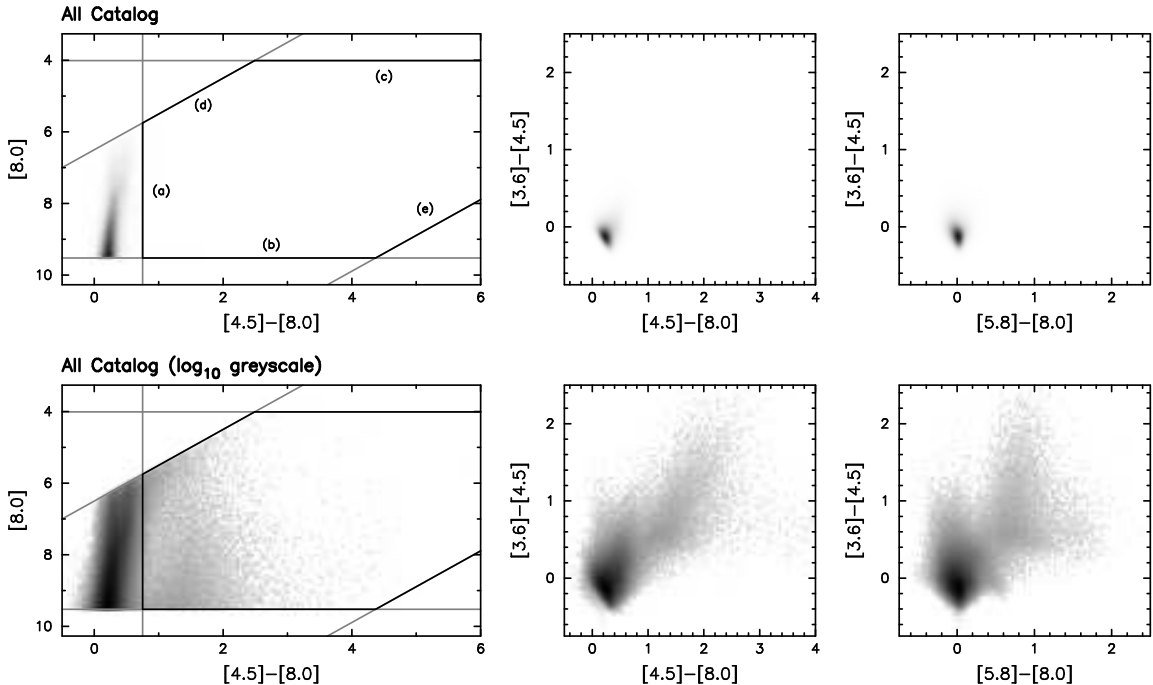}
\caption{All GLIMPSE~I and II Catalog sources satisfying equations (\ref{eq:brightness}) and (\ref{eq:quality}) shown on a linear grayscale (\textit{top}), and the same sources shown on a logarithmic grayscale (\textit{bottom}). The gray lines in the left panels show (as labelled in the top left panel): (a) a $[4.5]-[8.0]>0.75$ color selection, (b) the $[8.0]$ sensitivity limit, (c) the $[8.0]$ saturation limit, (d) the $[4.5]$ saturation limit, and (e) the $[4.5]$ sensitivity limit. The black lines show the boundaries of the resulting completeness region.\label{fig:selection}}
\end{center}
\end{figure}

In this section, only the sources that satisfy Equations (\ref{eq:brightness}) and (\ref{eq:quality}), i.e. the brightness and quality selection criteria, are used. Figure~\ref{fig:selection} shows a $[8.0]$ vs. $[4.5]-[8.0]$ color-magnitude diagram for all these sources (on a linear and logarithmic grayscale respectively). The large majority of sources lie close to, but not at $[4.5]-[8.0]=0$. This is likely due to the CO fundamental absorption feature in the spectrum of giants and dwarfs which decreases the 4.5\microns flux. Also shown are the distribution of sources in a $[3.6]-[4.5]$ vs. $[4.5]-[8.0]$ and a $[3.6]-[4.5]$ vs. $[5.8]-[8.0]$ color-color diagram. Most sources fall around $[3.6]-[4.5]\approx-0.1$, $[4.5]-[8.0]\approx+0.2$, and $[5.8]-[8.0]=0$. The slightly negative values for the $[3.6]-[4.5]$ color, the slightly positive value for $[4.5]-[8.0]$, and the zero value for $[5.8]-[8.0]$ are all consistent with the presence of an absorption feature at 4.5\microns.

\begin{figure}[p]
\begin{center}
\includegraphics[width=6.0in]{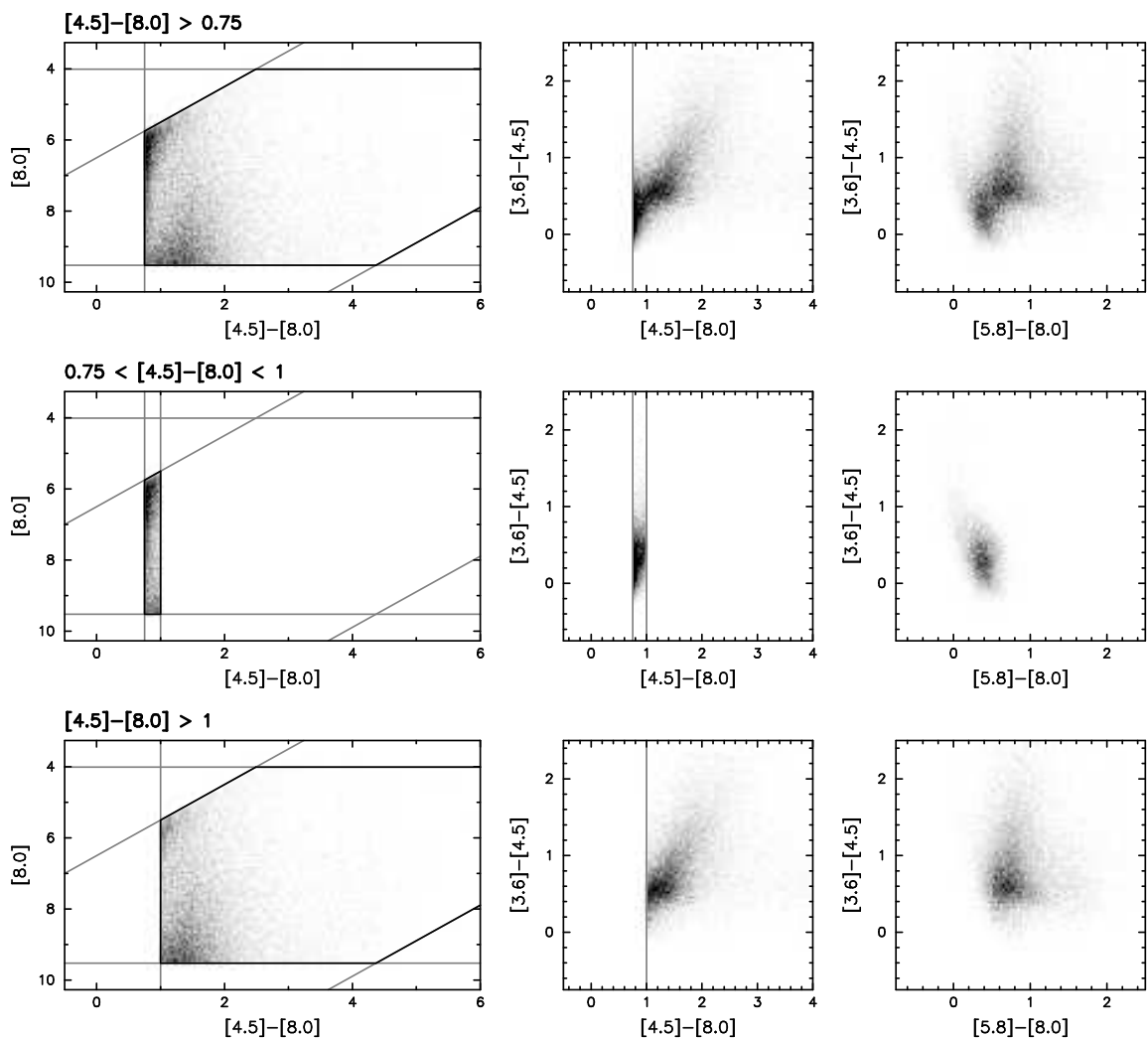}
\caption{As for Figure \ref{fig:selection}, but showing only sources with $[4.5]-[8.0]>0.75$ (\textit{top}), only sources with $0.75<[4.5]-[8.0]<1$ (\textit{center}), and only sources with $[4.5]-[8.0]\ge1$ (\textit{bottom}). All panels show the density of sources on a linear grayscale. The gray vertical lines in the central panels show the color selection criterion used in each case.\label{fig:selection2}}
\end{center}
\end{figure}

The top panel of Figure~\ref{fig:selection2} shows the sources selected if the selection criterion for red sources is chosen to be $[4.5]-[8.0]\ge0.75$. Two populations of sources are seen, the bluest of which contains the tail of the distribution of bright stars. The bottom two panels of Figure~\ref{fig:selection2}  show sources with $0.75\le[4.5]-[8.0]<1$ and $[4.5]-[8.0]\ge1$ respectively. The two populations separate well in color-color space. The $[3.6]-[4.5]$ vs. $[4.5]-[8.0]$ also clearly shows that one of the populations is in fact outliers from the main concentration of sources in color-color space.
For this reason, the final \textit{color} selection criterion for intrinsically red sources was chosen to be
\begin{equation}
\label{eq:color}
[4.5]-[8.0] \ge 1.
\end{equation}
This selection criterion corresponds to selecting all sources with a spectral index $\alpha\ge-1.2$  (as originally defined by \citealt{Lada:87:1}). Adopting the `Class' definition of \citet{Greene:94:614}, this means that all Class I sources ($\alpha \ge 0.3$), all `flat spectrum' sources ($-0.3 \le \alpha < 0.3$), and a large number of Class II sources ($-1.6 \le \alpha < -0.3$) will be included in the red source catalog. However, it should be stressed that the spectral index values discussed here are only calculated over a very small wavelength range.
The numbers of sources selected using this criterion are listed in Table~\ref{tab:glimpse}.
In total 22,099 sources were selected.

\subsection{Validation of the GLIMPSE photometry}

\label{sec:manual}

The quality selection criteria in Equation (\ref{eq:quality}) removed a large fraction of, but not all false red sources. For example, 8.0\microns fluxes can be over-estimated because of the spatially complex nature of the strong diffuse polycyclic aromatic hydrocarbon (PAH) emission - if a source lies on a sharp peak of diffuse emission, the background level can be under-estimated, and the flux over-estimated. Therefore, a foreground or background star superimposed on clumpy PAH emission can produce a false red star in the Catalogs.
In order to achieve close to 100\percent reliability, the following procedure was carried out:

\begin{enumerate}
\item For each red source from \S\ref{sec:red_selection}, the 4.5 and 8.0\microns fluxes were calculated independently using a custom written aperture and PSF photometry program, and using the final v2.0 mosaics rather than the BCD frames. For the sources extracted from the GLIMPSE~II first-epoch Catalogs, first-epoch only v2.0 mosaics were used. The point response functions (PRFs) were computed for the mosaics, as these differ slightly from BCD PRFs, and these were used to determine the appropriate IRAC aperture corrections, which were found to be in good agreement with the official SSC aperture corrections.
\item The mosaic images were examined by eye for every source, to determine whether the aperture photometry could be trusted, based on the radial profile of the source, by determining whether the sky background was correctly estimated, and whether there was any contamination inside the source aperture.
\item If the aperture photometry flux could not be trusted, a flux was determined by fitting the appropriate PSF to the source, and a residual image made by subtracting the scaled PSF from the mosaic. The residual image was inspected to determine whether this automated PSF photometry was reliable.
\item If the automated PSF photometry did not provide a clean residual, the position and flux of the scaled PSF were adjusted until the source was correctly subtracted.
\item In cases where this could not be achieved, the source was marked as unreliable. This includes extended sources, slightly diffuse sources, blended sources, and sources with low signal-to-noise detections.
\end{enumerate}

\begin{figure}[t]
\begin{center}
\includegraphics[width=6in]{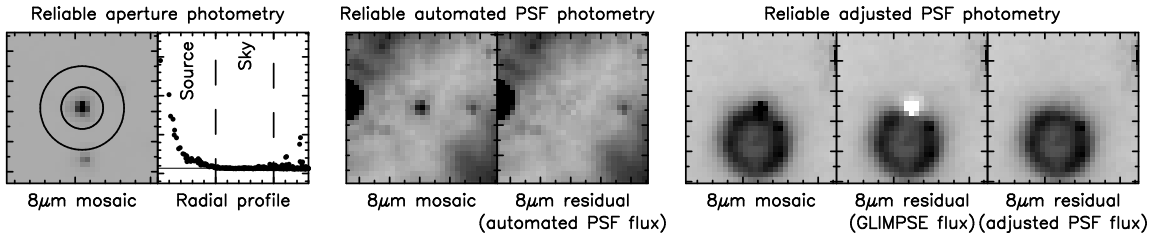}
\caption{Examples of methods used to derive independent mosaic photometry fluxes. \textit{Left:} a source for which a flux could be reliably measured using aperture photometry. The source aperture and the annulus for the sky determination are overplotted on the image. The radial profile shows that the sky background (indicated by a horizontal solid line) is accurately determined, and the source aperture (indicated by the left vertical dashed line) contains no contamination. \textit{Center:} a source for which automated PSF photometry provided a reliable mosaic flux. \textit{Right:} a source for which both the GLIMPSE and the independently determined flux over-estimated the flux of the source, but where the flux was reliably determined after adjusting it to obtain a clean residual. \label{fig:phot_methods}}
\end{center}
\end{figure}

Figure~\ref{fig:phot_methods} shows examples of the independent flux determinations for three sources - the first is one for which aperture photometry was reliable based on the radial profile of the source, the second is one for which the independent automated PSF photometry was reliable, and the third is one for which GLIMPSE's and the independent automated photometry were not reliable, but for which a reliable independent flux was obtained after manual adjustments. The independent photometry is likely to be on average more reliable than the GLIMPSE Catalog fluxes, not because of the flux determination algorithms, but because each source was visually inspected to decide which method produced the most reliable flux, and in some cases \textit{manual} adjustments were made to improve the PSF fit. In addition, using the mosaics rather than the original BCD frames means that outlying pixels such as cosmic rays do not affect the photometry as they were removed in the mosaicking process.

For sources that had either a reliable independent aperture or PSF (automated or adjusted) flux, these fluxes were compared with the original GLIMPSE fluxes, and all GLIMPSE fluxes that differed from the independent value by less than 15\percent of the GLIMPSE flux were considered to be trustworthy. For sources with fluxes disagreeing by a larger fraction, such as the source shown in the right-hand panel of Figure~\ref{fig:phot_methods}, it was found that the independently determined fluxes were always the most reliable, which is expected since their photometry was verified, and in some cases improved, manually. In total, only 1,098 and 302 sources (5.0\percent and 1.4\percent of all red sources respectively) had 4.5\microns and 8.0\microns fluxes respectively for which the GLIMPSE flux differed from the independent mosaic flux by more than 15\percent. In these cases, the independent flux was used instead of the original GLIMPSE flux. The brightness and color selection criteria from equations (\ref{eq:brightness}) and (\ref{eq:color}) were then re-applied.

Of the 22,099 red sources selected in \S\ref{sec:red_selection}, 3,055 (13.8\percent) were rejected because reliable independent photometry could not be performed in one or both of the bands (1,050 at 4.5\microns only, 1,171 at 8.0\microns only, and 834 at both 4.5 and 8.0\microns). Of those with reliable photometry, 58 were rejected as no longer red, and 59 were rejected as no longer bright enough or too bright (including 22 both no longer red and no longer bright enough or too bright). The source shown in the right hand panel of Figure~\ref{fig:phot_methods} is an example of a source that was only present in the original red source list because the 8.0\microns flux was overestimated - after computing reliable fluxes, it was found to have $[4.5]-[8.0]=0.38$ instead of $1.41$. Finally, 18,949 sources (85.7\percent) had reliable photometry, and satisfied the brightness and color selection criteria from equations (\ref{eq:brightness}) and (\ref{eq:color}), and these are the sources that constitute our final red source catalog; these sources are listed in Table~\ref{tab:red}.

\subsection{Constructing SEDs}

By default, the red source catalog contains  JH\ks, 3.6, and 5.8\microns data when available, since these are all in the GLIMPSE Catalogs, although none of these were \textit{required} for the selection. Of the 18,949 sources in the red source catalog, 13,011 have a \ks-band flux available, 9,740 have an H-band flux available, and 6,817 have a J-band flux available. Therefore, any JH\ks diagrams in subsequent sections will be incomplete as they only contain approximately one third of all sources.  The JH\ks, 3.6, and 5.8\microns photometry was not manually checked, but to ensure the high reliability of these fluxes, 3.6 and 5.8\microns fluxes were rejected if they did not satisfy the stringent quality criteria from equation (\ref{eq:quality}), and 2MASS fluxes were rejected if the quality flags in the 2MASS catalog were set to E, F, or X (indicating unreliable fluxes).

To determine MIPS 24\microns fluxes, PSF photometry was performed at the positions of the 18,949 red sources on the MIPSGAL 24\microns PBCD mosaics. Similarly to the independent IRAC fluxes, a custom-written PSF photometry program that allows adjustments in flux and position was used to manually improve the PSF fit for each source. Therefore, the MIPS photometry should also be reliable, modulo the uncertainties resulting from carrying out photometry on PBCD frames.

For sources saturated in the MIPS 24\microns observations, the MIPS 24\microns images were used to assess whether the MSX 21.3\microns flux would suffer from contamination from other sources. In cases where the MSX 21.3\microns emission originated solely from a single source, we list the MSX 21.3\microns flux in Table \ref{tab:red} instead of the MIPS 24\microns flux. We inspected the MSX images to remove any unreliable MSX 21.3\microns fluxes. For sources saturated at MIPS 24\microns where a reliable MSX 21.3\microns flux could not be used, we performed PSF photometry on the MIPS 24\microns data by fitting the unsaturated wings of the PSF whenever possible. In total, MIPS 24\microns fluxes were measured for 16,480 sources, MSX 21.3\microns fluxes were used for 112 sources, and 2,181 sources were either not covered by MIPS observations, not detected, were saturated, or could not have fluxes reliably measured (e.g low signal-to-noise, artifacts, blending), and the MSX 21.3\microns flux could not be used.

In the remainder of this paper, sources with MSX~21.3\microns data points instead of MIPS 24\microns are shown in color-color and color-magnitude plots as if they were MIPS 24\microns data points. As shown in \citet[][right panel of Figure 2]{Robitaille:07:2099}, the MIPS 24\microns to MSX 21.3\microns flux ratio is most likely to be between 0.7 and 3 for YSOs, meaning that by using the MSX 21.3\microns instead of MIPS 24\microns data in color-color or color-magnitude diagrams, the MIPS 24\microns magnitude is likely to be at most overestimated by 0.3\mag and underestimated by 1.2\mag. Compared to the range in $[8.0]-[24.0]$ shown in color-color diagrams in the remainder of this paper, and considering the small fraction of sources for which MSX 21.3\microns was used in place of MIPS 24\microns ($<$1\percent), this substitution is unlikely to be noticeable.

\subsection{Extended sources}

While the red source catalog presented in this paper aims to be as complete as possible within the color and magnitude/flux selection criteria imposed, it can only be complete for point sources, as it does not include YSOs that may be extended. While these are clearly of great importance, the GLIMPSE point source Catalog fluxes are measured using PSF photometry, and therefore are only reliable for point sources. Furthermore, extended sources have a lower probability of being found by the point source detection algorithm. For this reason, only a fraction of extended sources are likely to make it to the point source Catalogs, and in cases where they do, their fluxes are not likely to be reliable. Therefore, all extended sources were removed from the red source catalog for consistency, as described in \S\ref{sec:manual}.

In particular, massive YSOs that appear extended at 4.5\microns due to H$_2$ and CO bandhead emission from outflows are likely to be excluded; however, a number of these are identified in the recent work by \citet{Cyganowski:08} who found over 300 such objects in the GLIMPSE~I survey. In addition, the brightest extended YSOs at IRAC wavelengths may be detected as point sources in the MIPSGAL or MSX surveys, and may therefore be present in the MIPSGAL or MSX point source catalogs (such as the massive YSO candidates identified by the RMS survey).

\subsection{Completeness and Reliability}

In this section, we present estimates of the completeness and reliability of our red source catalog. The completeness - i.e. the fraction of red point sources in the survey area that are present in the final catalog - is not straightforward to estimate, as it depends on several factors, such as the initial completeness of the GLIMPSE Catalogs and the fraction of genuine sources rejected by the quality selection criteria in equation (\ref{eq:quality}). Rather than estimate the change in completeness at each step in the selection process, the completeness was estimated using sources from the GLIMPSE point source Archives, which are more complete (albeit less reliable) than the point source Catalogs used in this paper.
In total, 39,505 sources from the point source Archives satisfied equations (\ref{eq:brightness}) and (\ref{eq:color}). These were selected in the same way as the Catalog sources, i.e. using the GLIMPSE~II first epoch data where available. By examining a random sample of 400 of these sources and performing independent mosaic photometry, it was found that 27.0\percent were unambiguously not present in the mosaics, extended, not red enough, or too faint or bright. The remaining 73.0\percent included well-defined point sources, and sources for which no reliable photometry could be performed (e.g. blended sources or sources in areas of complex diffuse emission). Therefore, a conservative upper limit on the number of genuine red point sources was determined to be 39,505$\times$73.0\percent=28,839, assuming that the Archives themselves are complete. In comparison, 18,949 sources are present in our final red source catalog, suggesting a conservative lower limit on the completeness of 65.7\percent.

The reliability of the red source catalog, i.e. the fraction of sources that are genuine red point sources, should be virtually 100\percent, as all red sources in the catalog were examined individually to reject any sources that could not be confirmed as red. The 4.5\microns, 8.0\microns, and 24.0\microns photometry of all the red sources in the final catalog should be accurate to better than 15\%, since all fluxes were verified independently through mosaic photometry, and erroneous fluxes were corrected. For sources close to $[4.5]-[8.0]=1$, the photometric and calibration uncertainties mean that some sources may in fact have an intrinsic $[4.5]-[8.0]$ color slightly below $1$. However, a fraction of sources with a measured $[4.5]-[8.0]$ color slightly below $1$ are likely to have an intrinsic $[4.5]-[8.0]$ color slightly above $1$. The effect is likely to be similar in both directions, so that the total number of sources in our red source catalog should not be significantly affected by this. 

The most stringent criterion in the selection procedure used in this paper was the requirement that the 8.0\microns fluxes should be larger than 10\mJy. Although this was done in part to obtain a homogeneously sensitive sample of sources over the whole survey area, it also removed a large fraction of erroneous fluxes from faint and spurious sources, meaning that the subsequent quality selection criteria from equation (\ref{eq:quality}) only removed a further 10\percent of sources. However, the quality selection criteria would have been much more critical for fainter sources. As an example, approximately 40,000 of the sources in the entire GLIMPSE Catalog that have $[4.5]-[8.0]\ge0.75$ only have a single detection at 8.0\microns. The SEDs of these sources typically resemble `transition disk' SEDs, with photospheric fluxes for the first three IRAC bands, and an excess at 8.0\microns. However, upon closer inspection, it was clear that for the large majority of these sources, the color excess was not real, and in some cases the source was not present in the mosaic image at 8\microns. Most of these sources had 8.0\microns fluxes below 10\mJy, meaning that below this flux level, requiring two detections in each band would have been much more important.
In this light, we strongly recommend that when searching for red sources fainter than 10\mJy at 8\microns in the GLIMPSE~I or II Catalogs, one should make a careful assessment of the quality of the data and of the photometry. In particular, any source in the GLIMPSE Catalog with an SED resembling that of `debris' or `transition disks', i.e. with only an excess at 8\microns should not be trusted unless at least two detections are present at 8\microns, and the images are inspected visually to confirm the source. 
As previously mentioned in \S\ref{sec:initial_selection}, erroneous photometry is only an issue here because our red color selection preferentially selects sources with unreliable photometry from the catalogs, due to their unusual colors; but we emphasize that the GLIMPSE Catalogs are overall very reliable.

\section{Analysis}

\label{sec:analysis}

\subsection{Observable properties of the red sources}

\subsubsection{Angular distribution of sources}

\label{sec:spatial}

\begin{figure}[t]
\begin{center}
\includegraphics[width=6in]{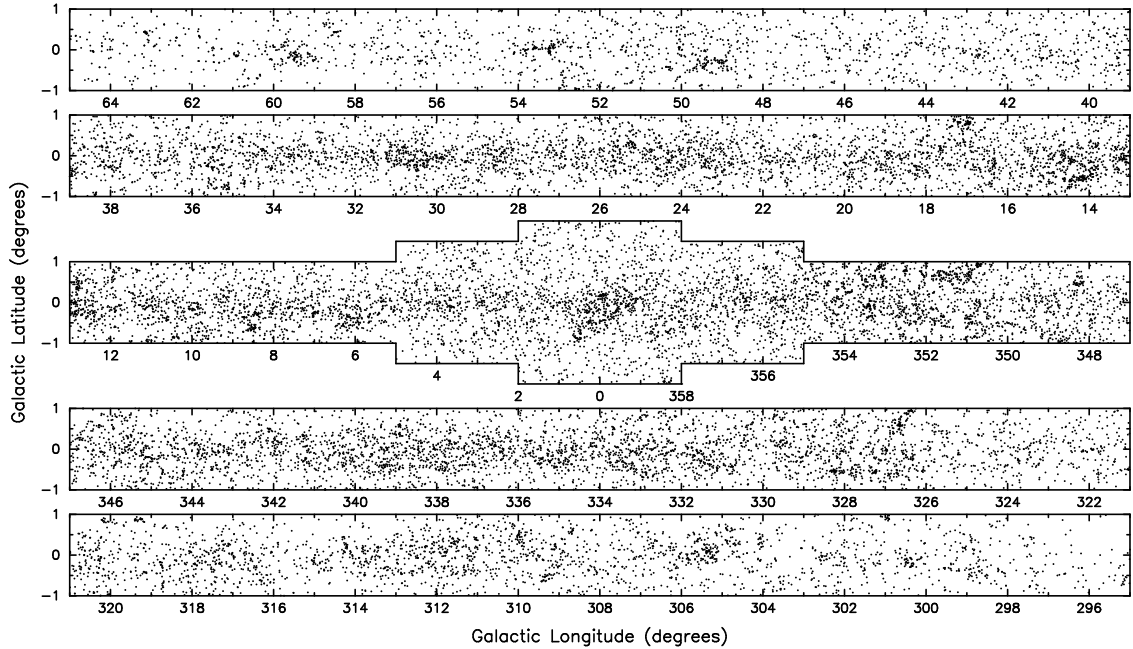}
\caption{Angular distribution of all the sources in the final red source catalog.\label{fig:spatial_red}}
\end{center}
\end{figure}

Figure~\ref{fig:spatial_red} shows the angular distribution of all intrinsically red sources in the final catalog. The distribution of sources shows a large number of clusters, and a more diffuse component that can be seen for example at latitudes $|b|>1$\degrees in the GLIMPSE~II region. The strong clustering of sources suggests that a significant fraction of the red sources are YSOs, as clustering is not expected for AGB stars or PNe, and galaxies are shown to contribute less than 0.5\% of the sources in the red source catalog (\S\ref{sec:galaxies}). The red sources do not show a strong increase in source density for $|\ell|<2$\degrees as was visible in the distribution of Catalog sources in Figure~\ref{fig:spatial_all}.

\subsubsection{Color and magnitude distribution}

\begin{figure}[t]
\begin{center}
\includegraphics[width=6in]{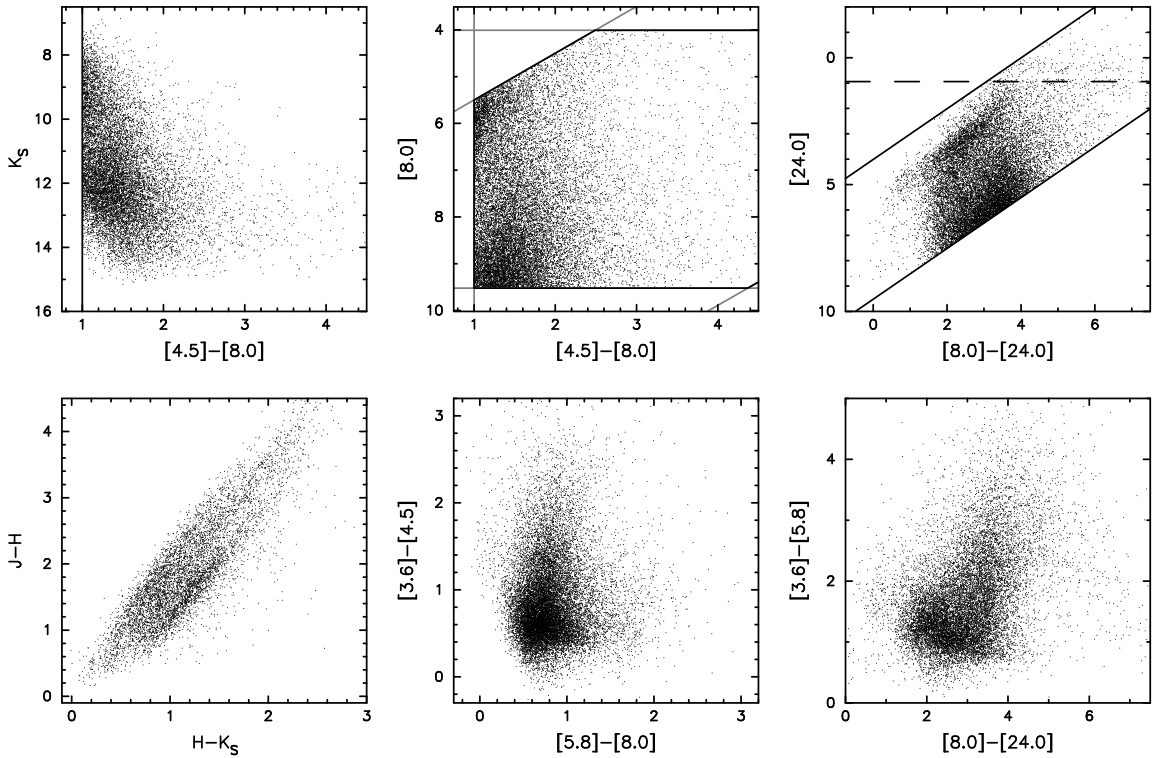}
\caption{Color/magnitude distribution of all the sources in the final red source catalog. The solid lines outline the selection criteria from equations (\ref{eq:brightness}) and (\ref{eq:color}). The horizontal dashed line in the top right panel shows the [24.0] magnitude limit above which MSX photometry is used if possible.\label{fig:color_red}}
\end{center}
\end{figure}

Figure~\ref{fig:color_red} shows the color-color and color-magnitude distribution of the red sources. It is clear from these diagrams that the red sources span a large region of color-color and color-magnitude space. In the color-magnitude diagrams, the red sources appear to separate into two populations: one redder and fainter, peaking at $[8.0]>8$ and $1<[4.5]-[8.0]<2$, and one bluer and brighter, peaking at $[8.0]<6.5$ and $[4.5]-[8.0]<1.4$. The latter appears to be an extension of the sources removed in \S\ref{sec:red_selection}. These two populations are identified in \S\ref{sec:populations} and separated in \S\ref{sec:decontamination}.

\subsubsection{Variability}

\label{sec:variability}

\begin{figure}[p]
\begin{center}
\includegraphics[width=6.0in]{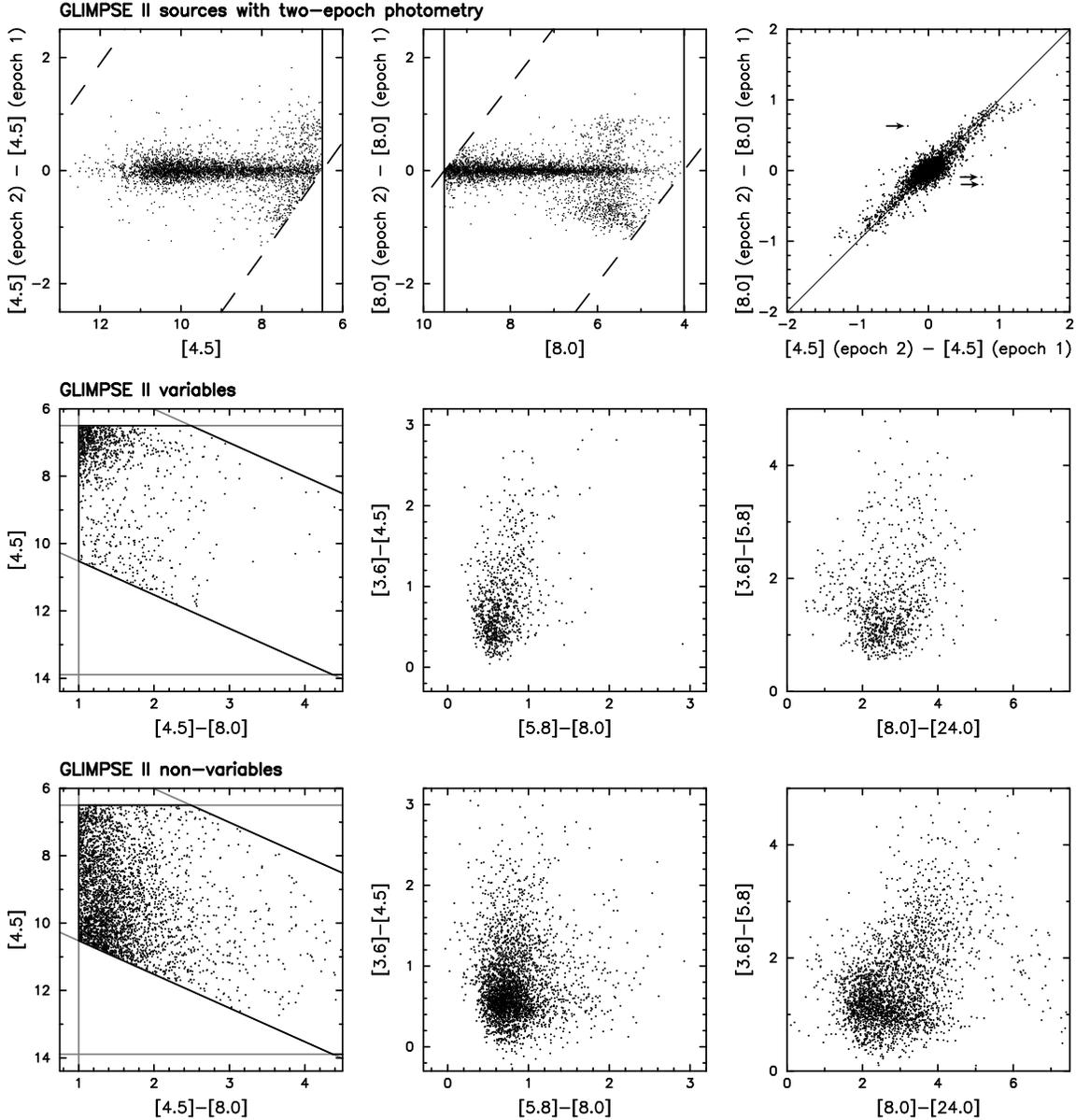}
\caption{\footnotesize \textit{Top} - Magnitude changes between the GLIMPSE~II epoch 1 and 2 data for the red sources where observations were available at both epochs. The solid and dashed lines in the top left and top center panels show the [4.5] and [8.0] sensitivity and saturation limits, and the solid diagonal line in the top right panel shows where the change in magnitude in equal in [4.5] and [8.0]. Most sources do not appear to change color significantly, i.e. the magnitude change is approximately equal in both bands. However, there are exceptions to this: for example for three extreme outliers from the distribution indicated by the arrows, the color change is so strong that it is clearly visible in 3-color GLIMPSE images. The cause of this is not clear, but this could simply be due to blending of a variable source with a foreground or background source. \textit{Center} - Color/magnitude distribution of the sources which show variability by at least 0.3\mag in one or both bands. The lines are as in Figure \ref{fig:color_red}. \textit{Bottom} - Color/magnitude distribution of the sources which vary by less than 0.3\mag in both bands between the two epochs .\label{fig:color_var}}
\end{center}
\end{figure}

As shown in Table~\ref{tab:glimpse}, 4,987 of the initially selected red sources were extracted from the GLIMPSE~II first epoch catalog. Of these, 4,472 are present in the final red source catalog after removal of unreliable sources, and 4,455 also fall in the GLIMPSE~II second epoch observations. The GLIMPSE~II second epoch point source Archives were used to assign second epoch fluxes\footnote{No second epoch point source Catalogs exist at the time of writing, because the current GLIMPSE Catalogs rely on the fact that each position is observed at least twice to increase the point source detection reliability, but GLIMPSE~II second epoch observations only consist of one observation for each position}. The same procedure as described in \S\ref{sec:manual} was carried out to ensure that these fluxes were reliable, namely computing independent mosaic fluxes and visually inspecting all sources to determine whether the second epoch fluxes were reliable. Reliable second epoch fluxes were available for 3,980 sources at 4.5\microns and for 4,331 sources at 8.0\microns. Most of the sources flagged as unreliable at 4.5\microns were above the saturation limit. The 4,455 sources with two-epoch photometry are listed in Table~\ref{tab:variables}.

The top panels in Figure~\ref{fig:color_var} show the change in [4.5] between the two epochs, the change in [8.0], and the correlation between the change in [4.5] and [8.0]. In total, 1,004 sources (22.5\percent of GLIMPSE~II sources with photometry at both epochs) show a change of at least 0.3\mag at either (or both) 4.5\microns and 8.0\microns.
At 4.5\microns, 308 become fainter, 298 brighter, and 398 do not have a reliable second epoch flux; at 8.0\microns, 242 sources become fainter, 586 brighter, and 176 do not have a reliable second epoch flux. The large number of unreliable fluxes at 4.5\microns is due to sources that brighten above the saturation limit in the second epoch. For sources that do have reliable fluxes at both wavelengths and epochs, the top right panel of Figure~\ref{fig:color_var} shows that in most cases the change in magnitude between the two bands is equal, and that the $[4.5]-[8.0]$ color does not change in most cases - although there are a few outliers.

Thus, it appears that approximately $2/3$ of the variable sources increase in brightness between the first and second epoch, of which half saturate at 4.5\microns in the second epoch, while only $1/3$ of sources become fainter. This clear asymmetry is a selection artifact rather than a physical effect: the variable sources tend to be close to the 4.5\microns saturation limit - therefore sources that would have become fainter in the second epoch are more likely to have been saturated during the first epoch, and therefore are less likely to be present in the red source catalog.

The colors and magnitudes of the variables and non-variables are shown in the middle and bottom panels of Figure~\ref{fig:color_var}. The variable sources are clearly not simply a random subset of all the red sources. Instead, it appears that the variable sources represent an important fraction of the population of bluer and brighter sources seen previously in Figure~\ref{fig:color_red}. As will be discussed further in \S\ref{sec:populations}, these sources are likely to be AGB stars that are Long Period Variables (LPVs). As shown by \citet{Marengo:08:331}, AGB stars with semiregular-type variability do not have IRAC amplitudes as large as AGB stars with Mira variability, and therefore the variable sources presented here are most likely to be Mira variables. We note that a fraction of variable stars may be mistakenly classified as non-variable if they happen to be at a similar light-curve phase at the two epochs, and therefore the fraction of variable sources should be considered a lower limit.

\subsection{Populations}

\label{sec:populations}

In this section, we attempt to qualitatively and quantitatively determine the composition of the red source catalog, which is expected to contain YSOs, AGB stars, PNe, and background galaxies.

\subsubsection{Planetary Nebulae}

\label{sec:pne}

The colors of PNe - that is, including both the central source and the diffuse emission - are known to be very red at IRAC and MIPS wavelengths, mainly due to varying contributions from H$_2$, PAH, dust, and ionized gas line emission, at 8\microns \citep{Hora:04:296,Hora:08:726}. Therefore, distant PNe are expected to be present in the red source catalog. Resolved PNe will not be present, as extended sources are excluded.

\begin{figure}[t]
\begin{center}
\includegraphics[width=6in]{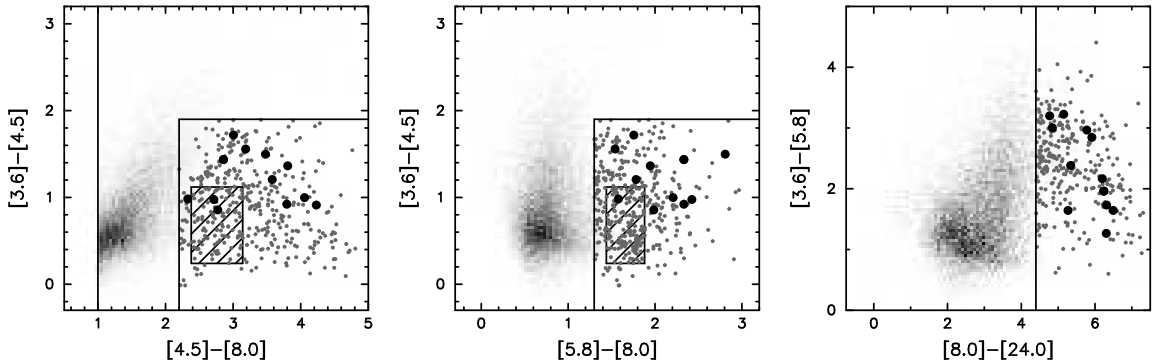}
\caption{IRAC and MIPS 24\microns color-color diagrams of all the red sources (grayscale), those that are known PNe (filled circles), and those that have similar colors to known PNe (dark gray points). The hatched rectangle shows the location of the Galactic PNe from \citet{Cohen:07:343}, and the solid lines show the selection criteria to select sources with colors similar to those of the known PNe. \label{fig:pne}}
\end{center}
\end{figure}

The red source catalog was cross-correlated with the Strasbourg-ESO Catalogue of Galactic Planetary Nebulae \citep{Acker:92}, the sample of southern PNe from \citet{Kimeswenger:01:115}, and the sources in the RMS survey confirmed as PNe \citep{Hoare:04:156}, which contain 1143, 995, and 76 sources respectively. Of these, only 2, 5, and 6 are present in the final red source catalog respectively. Figure~\ref{fig:pne} shows the location of these PNe in color-magnitude and color-color space relative to all the sources in the red source catalog. The PNe are clearly amongst the reddest sources in $[4.5]-[8.0]$, $[5.8]-[8.0]$ and $[8.0]-[24.0]$. The colors in $[3.6]-[4.5]$ vs. $[5.8]-[8.0]$ are in good agreement with the colors of PNe reported by \citet{Hora:04:296}, \citet{Cohen:07:343}, and \citet{Hora:08:726}. 

It is likely that previously unknown distant PNe are missing from the three catalogs used for the cross-correlation. An upper limit on the number of PNe in the red source catalog can be estimated by counting all red sources that fall within the same region of color-color space as the known PNe. To do this, we select all sources satisfying:
\begin{equation}
\left\{
\begin{array}{lcl}
\left[3.6\right]-\left[4.5\right] &<& 1.9\\
\left[4.5\right]-\left[8.0\right] &>& 2.2 \\
\left[5.8\right]-\left[8.0\right] &>& 1.3\\
\left[8.0\right]-\left[24.0\right] &>& 4.4\\
\end{array}
\right.
\end{equation}
For any given source, only the selection criteria that could be applied based on the data available were used: for example, for sources with no [3.6] magnitude, only the last three criteria were used. In total, 458 sources were selected. These selection criteria will also include objects which are not PNe (such as YSOs), but provide an \textit{upper limit} of 2.4\percent on the fraction of PNe in the red source catalog.

\subsubsection{Galaxies and AGNs}

\label{sec:galaxies}

\begin{figure}[t]
\begin{center}
\includegraphics[width=5.5in]{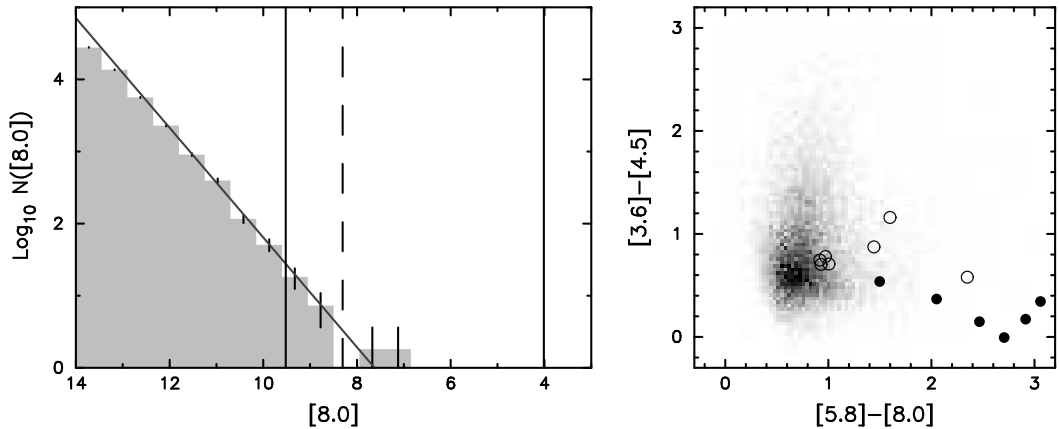}
\caption{\textit{Left:} the gray histogram shows the number density of SWIRE galaxies as a function of $[8.0]$ magnitude. The vertical solid lines show the range of magnitudes used to select the red sources in this paper, and the vertical dashed line shows the approximate SWIRE saturation level. The solid gray line shows the best fit to the source counts below the SWIRE saturation level. This line is used to extrapolate the expected number of galaxies above the saturation level. \textit{Right:} IRAC color-color diagram of all red sources (grayscale), and the SWIRE galaxies which would have satisfied the brightness criterion in the absence of interstellar extinction. The empty circles show AGN, which have dust-dominated SEDs at IRAC wavelengths, and the filled circles show `normal' galaxies, which have PAH-dominated SEDs in the same wavelength range. \label{fig:galaxies}}
\end{center}
\end{figure}

In order to estimate the fraction of galaxies and dusty AGNs present in the red source catalog, data from the \textit{Spitzer} Wide-Area Infrared Extragalactic Survey (SWIRE; \citealt{Lonsdale:03:897}) were used. These data consists of IRAC and MIPS observations of six patches of sky covering in total 63.2\sqdeg (at IRAC wavelengths). The Spring '05 Catalogs for the Lockman, ELAIS N1, ELAIS N2, and XMM\_LSS fields, and the Fall '05 Catalogs for the CDFS and ELAIS S1 fields were used.

Sources detected at both 4.5 and 8.0\microns were selected from the SWIRE Catalogs. Since our red source catalog only includes point sources, only sources with an extended source flag of -1, 0, or 1 at both these wavelengths (i.e. point-like, indeterminate, or slightly extended) were selected, and the IRAC aperture fluxes were used. Foreground stars were removed by selecting only sources with $[5.8]-[8.0]> 0.5$. The $[8.0]$ distribution of the sources satisfying these criteria is shown in Figure~\ref{fig:galaxies} along with the approximate saturation limit for SWIRE and the brightness criteria from Equation (\ref{eq:brightness}). Only 13 sources would have been bright enough to be selected, in the absence of interstellar extinction, of which 6 are known `normal' galaxies with PAH-dominated mid-infrared colors, and the remaining 7 are known active galaxies (specifically QSOs and Seyfert II galaxies), with dust-dominated mid-infrared colors. The IRAC colors of these sources are shown in Figure~\ref{fig:galaxies}.

Extrapolating the distribution of sources above the SWIRE saturation level (as shown in Figure~\ref{fig:galaxies}) suggests that approximately 16$\pm$4 sources might have been `detected' in total in the absence of saturation. The GLIMPSE survey area is 274\sqdeg, so the number of `detected' galaxies should be scaled accordingly, assuming that the density of extra-galactic sources in the six SWIRE patches is similar to that in the GLIMPSE survey area (in the absence of interstellar extinction). This results in an estimated $70\pm17$ galaxies in the red source catalog. This number is an upper limit, as for a large fraction of the GLIMPSE area, the extinction through the Galaxy is likely to make these sources too faint to be included in our red source catalog. Therefore, \textit{at most} 0.4\percent of sources in the red source catalog are likely to be background extragalactic sources. We note that this upper limit applies only to the red source catalog presented in this paper. The upper limit on the number or fraction of galaxies in the entire GLIMPSE Catalogs is likely to be different if for example the 8\microns flux is not required to be larger than 10\mJy.

\subsubsection{AGB stars}

\label{sec:agb}

\begin{figure}[tp]
\begin{center}
\includegraphics[width=6in]{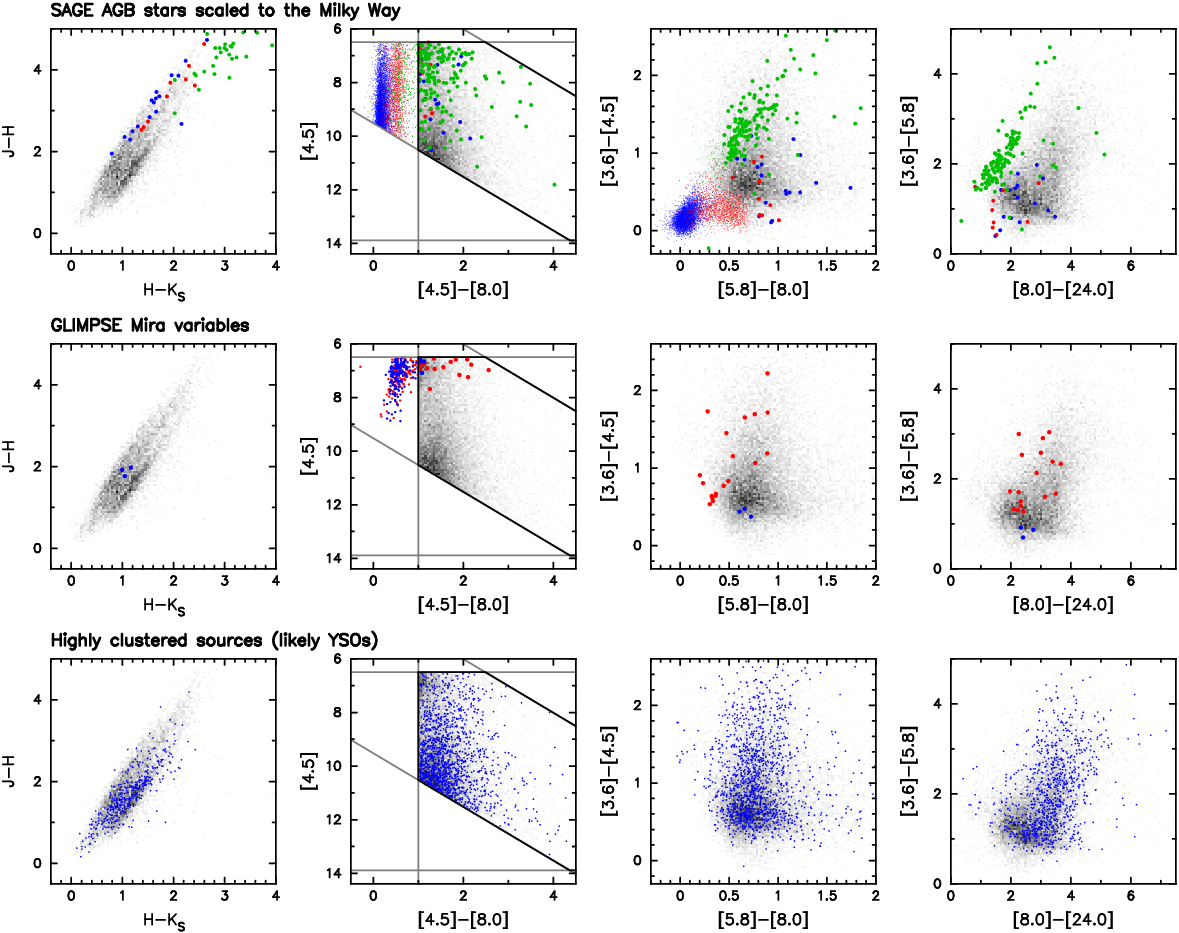}
\caption{Location of AGB stars and likely YSOs in IRAC and MIPS color-magnitude and color-color space. In all plots, the background grayscale shows all red sources, and the solid lines are as in Figure~\ref{fig:color_red}. \textit{Top:} location of the SAGE LMC AGB stars, scaled to random locations in the Milky-Way, that satisfy the brightness and color selection criteria from equations (\ref{eq:brightness}) and (\ref{eq:color}). C-rich \sagb stars are shown in red, O-rich \sagb stars in blue, and \xagb stars in green. The smaller points in the $[4.5]$ vs. $[4.5]-[8.0]$ and the $[3.6]-[4.5]$ vs. $[5.8]-[8.0]$ diagrams are the sources that do not satisfy the color selection criterion. \textit{Center:} Mira variables from the GCVS (red) and OGLE (blue), identified amongst the GLIMPSE sources that satisfy the brightness, quality, and color selection criteria. The smaller points in the $[4.5]$ vs. $[4.5]-[8.0]$ diagram are the sources that do not satisfy the color selection criterion. \textit{Bottom:} highly clustered red sources (blue). In all of the JH\ks plots, sources that are not detected in J, H, or \ks are not shown.\label{fig:agbyso}}
\end{center}
\end{figure}

In order to understand where various types of AGB stars should lie in IRAC and MIPS 24\microns color-color and color-magnitude space, the red source catalog was initially cross-correlated with catalogs of known C- and O-rich AGB stars \citep[e.g.][]{Lindqvist:92:43,Chengalur:93:189,Sjouwerman:98:35,Alksnis:01:1}. However, no previously known C-rich, and only 16 previously known O-rich AGB stars were found.  Instead, the colors and magnitudes of AGB stars were analyzed using known AGB stars in the `Surveying the Agents of a Galaxy's Evolution' (SAGE) survey of the Large Magellanic Cloud (LMC; \citealt{Meixner:06:2268}; PID 20203).

We used the SAGE IRAC and MIPS magnitudes for AGB stars from Srinivasan et. al (2008, in preparation).
These include `standard' C- and O-rich AGB stars as well as `extreme' AGB stars\footnote{We have included red sources that were excluded in their analysis (Srinivasan, private communication, 2008), and have removed 29 `extreme' AGB stars that were likely LMC YSOs following a visual inspection}. The latter, sometimes referred to as `obscured' AGB stars, are also C- and O- rich AGB stars, but with very high  mass-loss rates, and therefore large amounts of circumstellar dust. In the remainder of this paper, `Extreme' AGB stars will be referred to as `\xagb stars', and `Standard' AGB stars as `\sagb stars'. We will use the term `AGB stars' when referring to both \xagb and \sagb stars.

The LMC AGB stars were assumed to be situated at a distance of 50.1\kpc \citep[][and references therein]{Alves:04:659}. Using GLIMPSE data, \citet{Benjamin:05:L149} showed that red giant stars in the Milky-Way are distributed in an exponential disk with a scale length of 3.9\kpc. We made the assumption that the AGB stars are similarly distributed, and randomly sampled positions in such a disk, with an exponential vertical distribution that had a scaleheight of 300\pc. Using this 3-D distribution, 41\percent of AGB stars fell inside the GLIMPSE survey area (as defined in \S\ref{sec:irac_description}). The magnitudes were scaled appropriately, and interstellar extinction was applied using an approximate extinction-distance relation of 1.9\,mag/kpc \citep{Allen:73}, using the extinction law derived by \citet{Indebetouw:05:931} for JH\ks and IRAC wavelengths, and assuming A$_{24\mu{\rm m}}$/A$_{\rm V}\sim$0.04, by extrapolation of the mid-infrared interstellar extinction law found by \cite{Lutz:99:623}. We then selected only sources that matched the selection criteria described in \S\ref{sec:initial_selection}. 

The SAGE sensitivity limits (17.47 and 14.23 at 4.5 and 8.0\microns respectively) are 3-4 magnitudes lower than the flux requirements from \S\ref{sec:initial_selection}, so that the faintest SAGE sources were still fainter than our sensitivity requirements after placing the synthetic AGB stars at positions in the Milky-Way and applying extinction, except for the few sources that were placed at distances very close to the Sun. Therefore, this method should be sensitive to AGB stars in all of the color-magnitude space covered by the red source catalog. For the remainder of this section, these re-scaled and reddened LMC AGB stars will be referred to as the \textit{synthetic} AGB stars. The main caveat of this approach is that it is not clear whether the intrinsic IRAC and MIPS 24\microns colors of AGB stars are identical in the LMC and the Milky-Way. However, any differences between the intrinsic colors between the LMC and Milky-Way are likely to be much less important than the uncertainty in the extinction-distance relation and in the extinction law assumed.

The location of the synthetic AGB stars that fall in the survey area in color-magnitude and color-color space is shown in Figure \ref{fig:agbyso}: just under half (46\percent) satisfy the brightness selection criteria in equation (\ref{eq:brightness}), of which 72\percent are O-rich \sagb stars, 24\percent are C-rich \sagb stars, and 4\percent are \xagb stars. Under 1\percent of all AGB stars that fall in the survey area  and that satisfy the brightness selection criteria further satisfy the color selection criterion from equation (\ref{eq:color}). Of these, 83\percent are \xagb stars, 10\percent are O-rich \sagb stars, and 7\percent are C-rich \sagb stars. These account for 24\percent, 0.1\percent, and 0.4\percent of all \xagb, O-rich \sagb, and C-rich \sagb stars that fall inside the survey area respectively. Therefore, \xagb stars are likely to represent the majority of AGB stars in our red source catalog, as they are much redder on average than \sagb stars. The \xagb stars that are present in the catalog likely represent a quarter of all \xagb stars that fall in the survey area, while the \sagb stars only represent a very small fraction ($<0.5$\percent) of those in the survey area, because most are bluer than $[4.5]-[8.0]=1$.

Since \xagb stars are luminous, they are more likely to be present at the bright end of the red source sample. In fact, a large fraction of \xagb stars appear to share the same region of $[4.5]$ and $[8.0]$ color-magnitude space as the variable stars shown in Figure~\ref{fig:color_var}, which can be explained if the variable stars in this region are AGB stars with Mira variability. Evidence that this is the case is shown in the central panel of Figure~\ref{fig:agbyso}, which shows the location of 274 GLIMPSE Mira variables, identified amongst the GLIMPSE sources satisfying equations (\ref{eq:brightness}) and (\ref{eq:quality}) using the Combined General Catalog of Variable Stars v4.2 \citep{Samus:04} and the OGLE catalog of Mira variables in the Galactic Bulge \citep{Groenewegen:05:143}. Of these, 25 also satisfy the color selection from equation (\ref{eq:color}) and therefore are in the red source catalog. Only three of the 25 fall in the region covered by the two epoch GLIMPSE~II observations, but all three are classified as variable at 4.5 and 8.0\microns based on the analysis in \S\ref{sec:variability}.

In J$-$H vs. H$-$\ks color-color space, the synthetic \sagb stars tend to occupy bluer colors of H$-$\ks. The \xagb stars tend to have large J$-$H values, suggesting very high values of the interstellar extinction. In IRAC color-color space, the AGB stars are not easily distinguishable from the whole population of red sources. However, once MIPS 24\microns data are included, the AGB stars tend to have bluer $[8.0]-[24.0]$ colors on average than the overall red population (e.g. only very few synthetic AGB stars are seen to have $[8.0]-[24.0]>3$).

In order to roughly estimate the fraction of AGB stars in the red source catalog, it is necessary to know the ratio of the total number of AGB stars in the Milky-Way to the total number of AGB stars in the LMC. Assuming that the ratio of AGB to all stars is approximately the same for both galaxies, and that both have similar star formation histories, this ratio can be derived from the ratio of the total stellar mass in each galaxy (excluding the gas and dark matter mass). The total stellar mass of the Milky-Way is of the order of $4.8-5.5\times10^{10}$\,M$_\odot$ \citep{Flynn:06:1149}, and that of the LMC is of the order of $2.7\times10^9$\,M$_\odot$ \citep{Marel:02:2639}, implying a ratio in stellar masses and therefore in the number of AGB stars of approximately 20. Since 177 synthetic AGB stars are `selected' according to the brightness and color selection criteria in equations (\ref{eq:brightness}) and (\ref{eq:color}), this suggests that approximately $177\times20=3,540$ AGB stars may be present in the red source catalog (i.e. 19\percent of all red sources). 

\subsubsection{Young stellar objects}

\label{sec:yso}

In order to understand where YSOs lie in near- and mid-infrared color and magnitude space, sources that are highly likely to be YSOs need to be identified in the red source catalog. However, catalogs of well known Galactic YSOs, such as confirmed massive YSOs from the RMS survey, cannot be used, as these surveys are not as deep as GLIMPSE and would appear to show that YSOs are more likely to be present at the bright end of the red source catalog.

Instead, a method involving only the red sources presented in this paper was used. As described in \S\ref{sec:spatial}, a large number of clusters are present in the red source catalog. However, AGB stars and PNe are not expected to be clustered in this way, and galaxies represent less than 0.5\% of the sources (\S\ref{sec:galaxies}). Therefore clustered objects have a high probability of being YSOs. By requiring the second closest neighbor of a source to be less than 2\arcmin, only sources in the regions of highest source density were selected. These include for example a number of sources in the NGC 6611 cluster in the Eagle Nebula \cite[M16; e.g.][]{Indebetouw:07:321}.

The distribution of these sources in color-color and color-magnitude diagrams is shown in the bottom panel of Figure~\ref{fig:agbyso}. In the J$-$H vs. H$-$\ks diagram, the probable YSOs appear to occupy preferentially redder values of H$-$\ks. In the $[4.5]$ vs. $[4.5]-[8.0]$ color magnitude diagram, the distribution of probable YSOs matches that of the overall distribution of red sources, with the exception of the bright blue peak where AGB stars are expected to lie. In $[3.6]-[4.5]$ vs. $[5.8]-[8.0]$ color-color space, the clustered sources are virtually indistinguishable from the overall red source sample. Finally, in $[3.6]-[5.8]$ vs. $[8.0]-[24.0]$ color-color space, the clustered sources do not match the distribution of all the red sources, as there is a deficit of clustered sources for $[8.0]-[24.0]\lesssim2.5$.

\subsection{Separation of YSOs and AGB stars}

\label{sec:decontamination}

\begin{figure}[p]
\begin{center}
\includegraphics[width=6.0in]{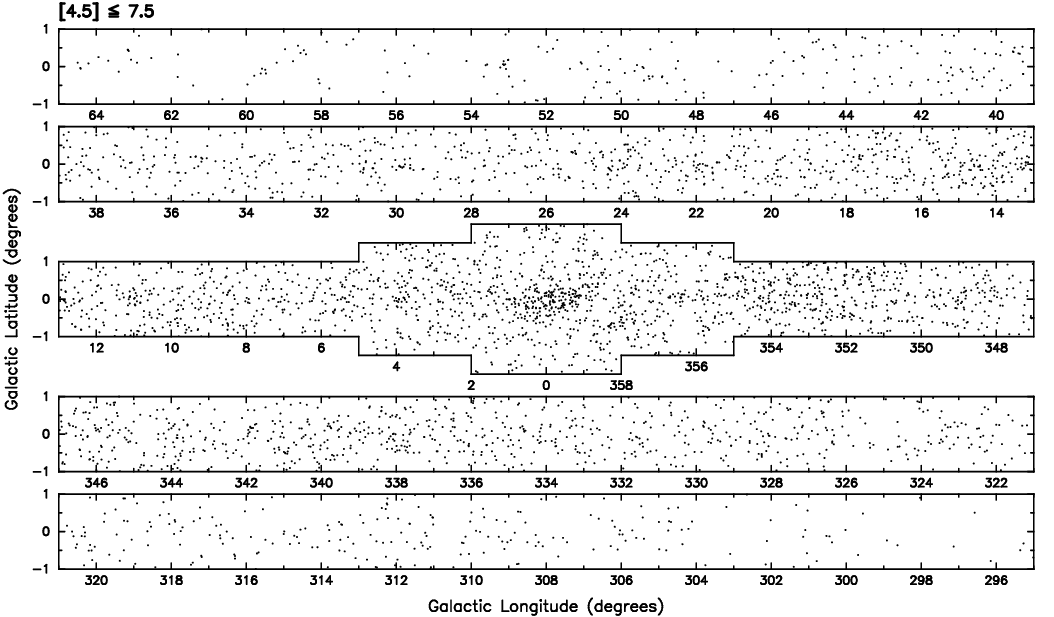}\\
\vspace{0.1in}
\includegraphics[width=6.0in]{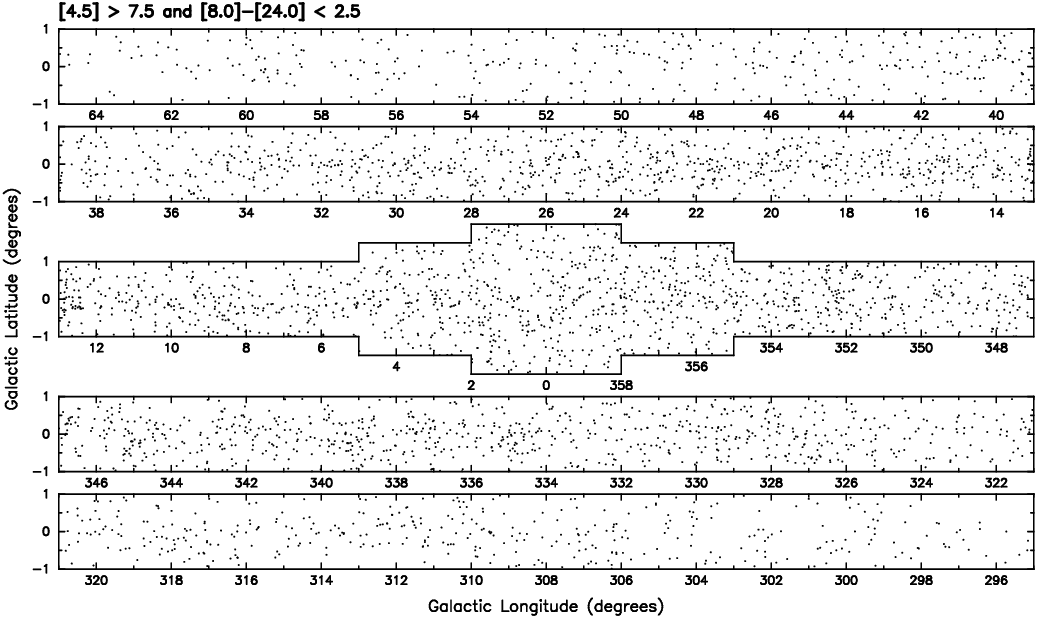}
\caption{Angular distribution of the color selected populations consisting mostly of \xagb stars (top), and \sagb stars (bottom).\label{fig:agb}}
\end{center}
\end{figure}

In this section, we present an estimate of the relative fraction of AGB stars and YSOs in the red source catalog, using simple color-magnitude selection criteria to separate the two populations as much as possible. We emphasize that the purpose of this analysis is not to provide reliable selection criteria for the two populations, but simply to estimate the relative importance of each population. Therefore, the separation is only approximate, and there is likely to be contamination in both directions. A reliable separation of the two populations would require additional data, such as spectroscopic observations, which are not available here.

As shown in Figure \ref{fig:agbyso}, \xagb stars account for the brighter and bluer peak seen in $[8.0]$ or $[4.5]$ vs. $[4.5]-[8.0]$ color-magnitude diagrams. Therefore, we classify all sources with $[4.5]\le7.8$ as \xagb star candidates, although we note that this criterion will inevitably include a fraction of luminous YSOs. The angular distribution of these sources is shown in the top panel of Figure \ref{fig:agb}, and does not show any clustering, suggesting that the fraction of YSOs  with $[4.5]\le7.8$ is small. The number density of sources increases towards the Galactic center, and a clear peak is seen at the Galactic center itself, suggesting that this population of sources is intrinsically very luminous, and is seen out to at least 8.5\kpc. In the GLIMPSE~II region, 52\percent of sources selected as \xagb stars are variable, and 68\percent of variable stars are \xagb stars.

\begin{figure}[p]
\begin{center}
\includegraphics[width=6.0in]{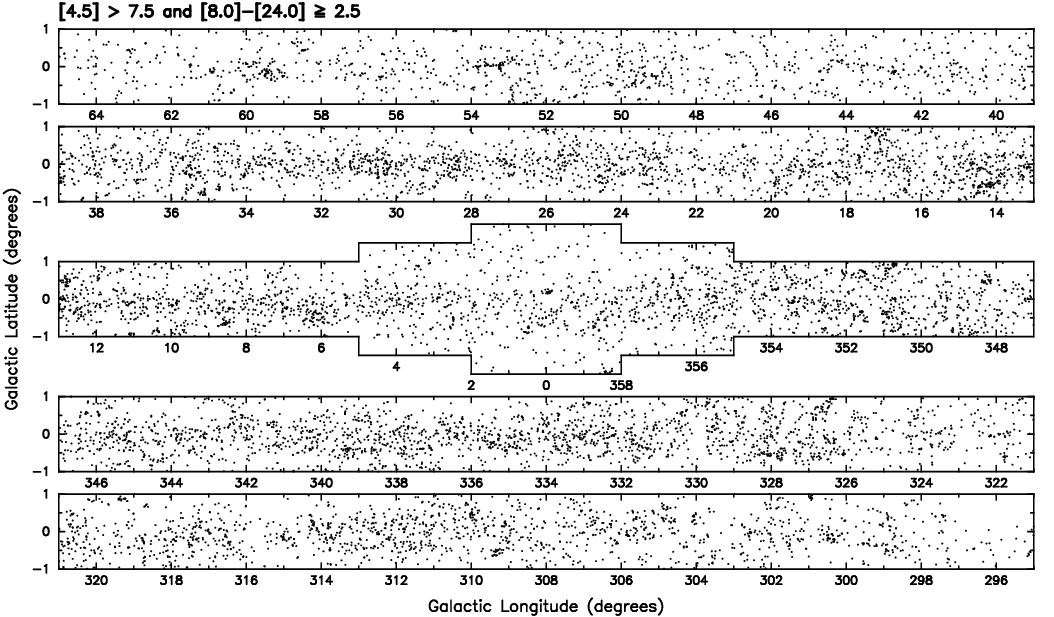}\\
\vspace{0.1in}
\includegraphics[width=6.0in]{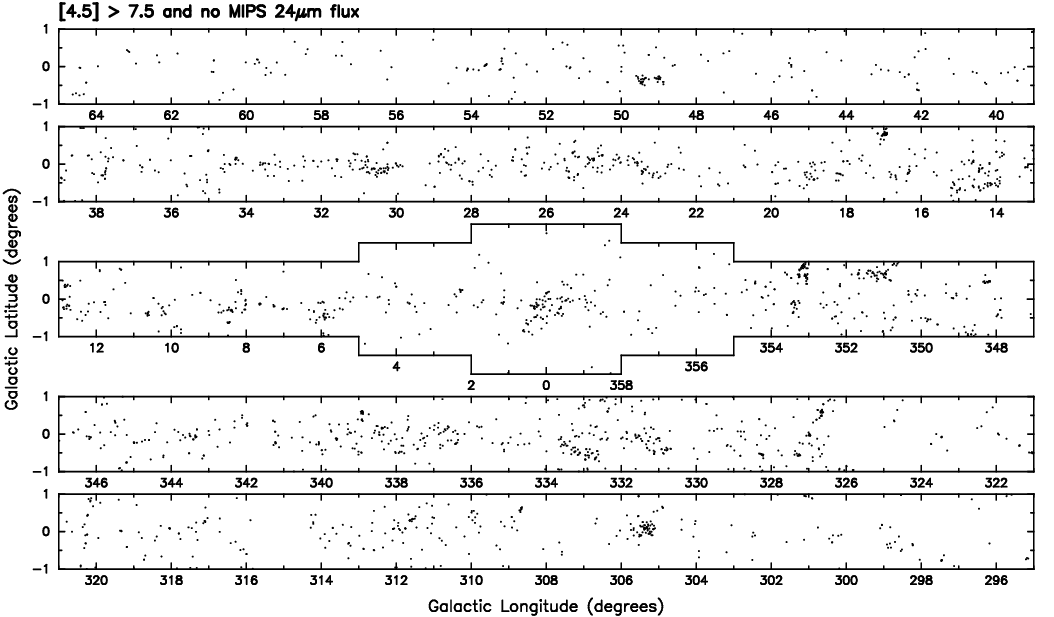}
\caption{Angular distribution of the color selected populations consisting mostly of YSOs (top), and the sources with no MIPS 24\microns photometry (bottom), which also consist mostly of YSOs.\label{fig:yso}}
\end{center}
\end{figure}

Sources with $[4.5]>7.8$ are likely to consist mostly of \sagb stars and YSOs. After removal of most of the \xagb stars as described above, both distributed and clustered sources remain. A small sample of the distributed sources was extracted by selecting sources seen at high latitudes ($|b|>1.5$\degrees) in the GLIMPSE~II region, where there is no indication of ongoing star formation. These are most likely to be \sagb stars. After examining the colors of these sources, it was apparent that a number of these had IRAC and MIPS colors consistent with the peak of sources seen in Figure \ref{fig:agbyso} for $[8.0]-[24.0]<2.5$, where there was also a deficit of YSO candidates (c.f. \S\ref{sec:yso}). Since the red sources were all required to have $[4.5]-[8.0]\ge1$ (equation [\ref{eq:color}]), sources with $[8.0]-[24.0]<1.9$ have SEDs with spectral indices that are lower between 8 and 24\microns than between 4.5 and 8\microns. This behavior is not typical of YSO SEDs, and 
it indicates that the majority of the dust lies very close to the photosphere of the central source, as is more likely to be the case for AGB stars. A number of the high latitude stars also appeared to have $1.9<[8.0]-[24.0]<2.5$. 
Therefore, a cutoff value of $[8.0]-[24.0]=2.5$ was chosen to separate candidate \sagb stars from candidate YSOs.
The angular distribution of red sources detected at 24\microns and with $[8.0]-[24.0]<2.5$ is shown in the bottom panel of Figure \ref{fig:agb}. These sources appear to be uniformly distributed in the Galactic plane, confirming that these are likely to be mostly \sagb stars. The number density of these sources falls of somewhat slower with Galactic longitude than the \xagb star candidates, consistent with their lower luminosity and consequently closer distance. In contrast, the distribution of red sources with $[8.0]-[24.0]\ge2.5$, shown in the top panel of Figure \ref{fig:yso}, shows more clustering, in agreement with their classification as candidate YSOs. The absence of a peak at the Galactic center for \sagb stars can be explained if these are not seen as far out as the Galactic center. However, while they are indeed fainter than \xagb stars, the analysis in \S\ref{sec:distances} shows that they should still be detectable beyond 8.5\kpc, so it is likely that this is instead an artifact due to the requirement for a valid MIPS 24\microns magnitude to carry out the $[8.0]-[24.0]$ color selection (see below).

This criterion used to separate \sagb stars from YSOs is similar to the $[8.0]-[24.0]>2.2$ criterion suggested by \citet[][equation (3)]{Whitney:08:18} for a stringent removal of AGB stars. We note that this simple color selection only approximately separates AGB stars and YSOs, so that there is likely to be contamination in both directions. For example, an examination of the IRAC and MIPS colors of spectroscopically confirmed AGB stars in Serpens suggests that the $[8.0]-[24.0]<2.5$ criterion does appear to be successful in selecting a large fraction of AGB stars, but inevitably also selects a small fraction of YSOs (Paul Harvey and Neal Evans, 2008, private communication). Nevertheless, as the angular distributions of the various populations show, this separation is likely to be sufficient for the purpose of estimating the relative fraction of AGB stars and YSOs in the red source catalog.

Sources with $[4.5]>7.8$ and with no fluxes at MIPS 24\microns cannot be separated into \sagb stars and YSOs as suggested above. These sources show very strong clustering, as shown in the bottom panel of Figure \ref{fig:yso}. The two main reasons that some red sources do not have fluxes at 24\microns are either that they lie on top of very bright (and in some cases saturated) diffuse emission, which lowers the point source sensitivity, or that the photometry could not be carried out due to blending of multiple sources, since MIPS 24\microns has a lower angular resolution than IRAC. High stellar densities and bright 24\microns diffuse emission are most likely to occur in massive star formation regions and towards the Galactic center. Thus, it seems likely that most of the clustered sources that do not have MIPS 24\microns fluxes are YSOs, with the exception of the concentration of sources at the Galactic center, which may be the missing peak of \sagb stars mentioned previously.

To summarize, sources with 
\begin{equation}
[4.5]\le7.8
\end{equation}
are classified as candidate \xagb stars, sources with
\begin{equation}
[4.5]>7.8~~{\rm and}~~[8.0]-[24.0]<2.5
\end{equation}
are classified as candidate \sagb stars, and sources with
\begin{equation}
[4.5]>7.8~~{\rm and}~~[8.0]-[24.0]\ge2.5.
\end{equation}
are classified as candidate YSOs. Sources with $[4.5]>7.8$ and no MIPS~24\microns detections cannot be separated using the above criteria, but are likely to be dominated by highly clustered YSOs in massive star formation regions. Therefore, for the remainder of this section, these are classified as candidate YSOs.

\begin{figure}[p]
\begin{center}
\includegraphics[width=6.0in]{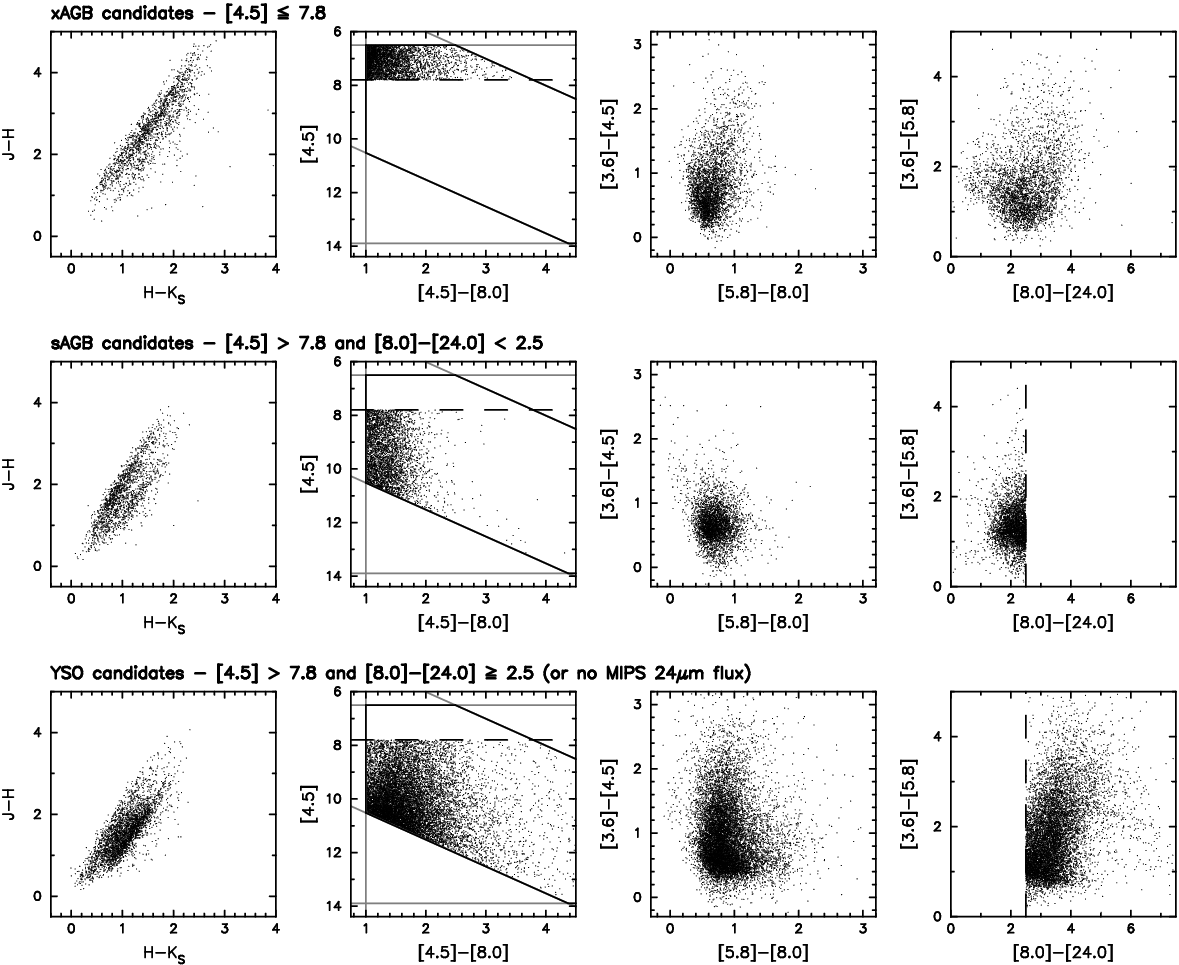}\\
\caption{Color-magnitude and color-color diagrams for the populations consisting mostly of \xagb stars (top), \sagb stars (center), and YSOs (bottom). The solid lines are as in Figure \ref{fig:color_red}. The dashed lines show the selection criteria used to separate the populations.\label{fig:color_decon}}
\end{center}
\end{figure}

Figure \ref{fig:color_decon} shows the color-color and color-magnitude distributions of the candidate \xagb stars, the candidate \sagb stars, and the candidate YSOs. The J$-$H vs. H$-$\ks diagrams show that the majority of sources in the three populations appear to have different near-infrared colors. The candidate \xagb stars have a larger extinction, consistent with the fact that these sources are on average more luminous than \sagb stars and are therefore seen out to larger distances. The candidate \sagb stars have a bluer H$-$\ks color than the candidate YSOs. In fact, the  JH\ks colors of these three populations are in good agreement with those of the synthetic \xagb stars, \sagb stars, and the clustered YSOs shown in Figure \ref{fig:agbyso}. The $[4.5]$ magnitude distribution of the \sagb stars and YSOs also differs: these show a quasi-uniform distribution of sources as a function of $[4.5]$, while the candidate YSOs show a clear increase in the number of sources towards fainter values of $[4.5]$. Finally, the distributions of the three populations in IRAC color-color space overlap, but are nevertheless distinctly different.

\begin{figure}[p]
\begin{center}
\includegraphics[width=6.0in]{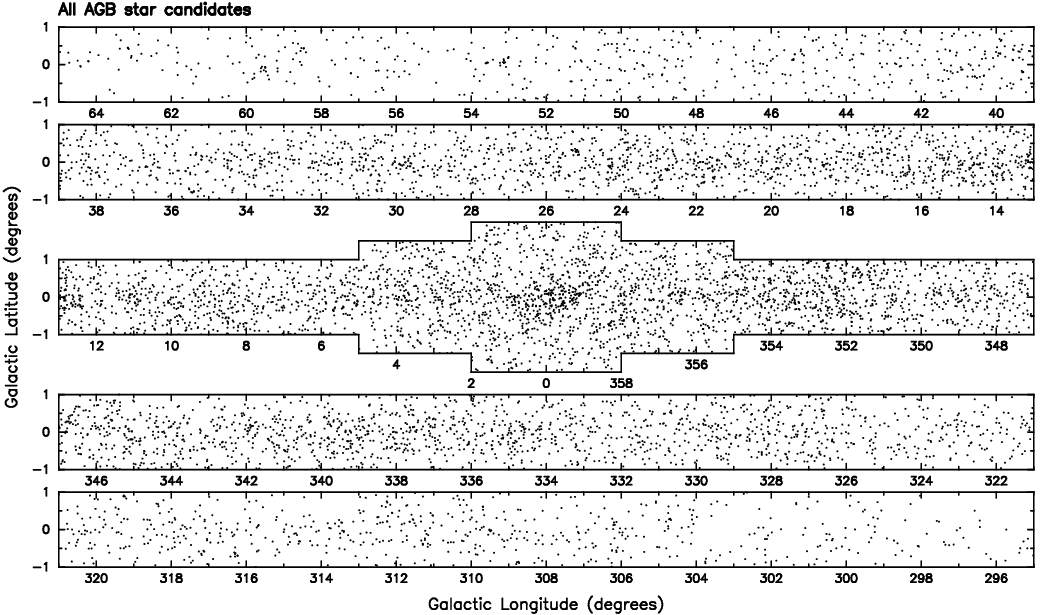}\\
\vspace{0.1in}
\includegraphics[width=6.0in]{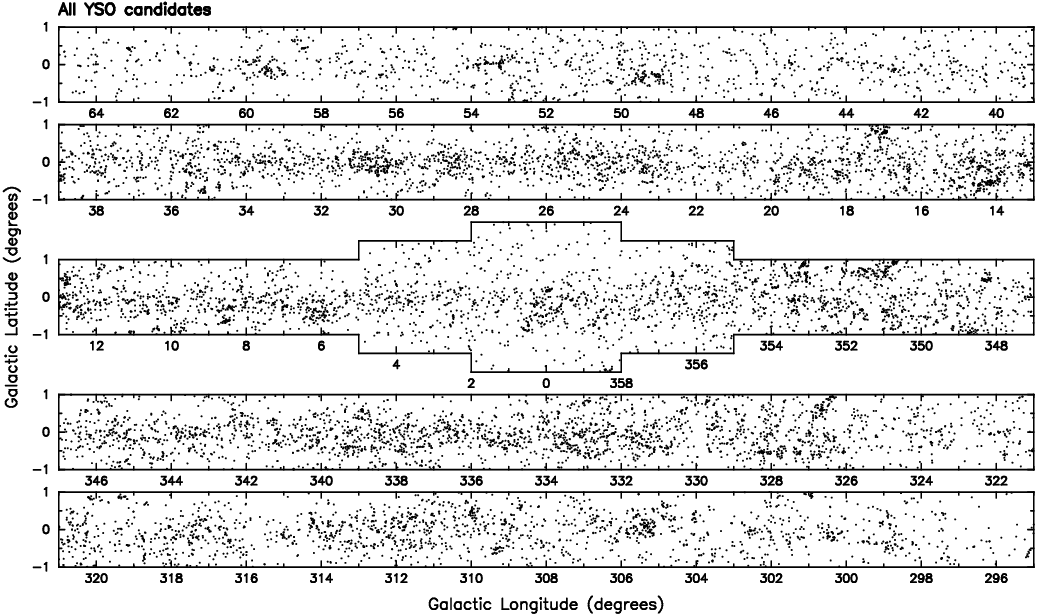}
\caption{Angular distribution of all the AGB candidates (top) and YSO candidates (bottom).\label{fig:separated}}
\end{center}
\end{figure}

Both the angular distribution and the colors of the various populations suggest that these are indeed mostly composed of \xagb stars, \sagb stars, and YSOs. In total, 7,300 and 11,649 sources are classified as candidate AGB stars and YSOs respectively, although we stress that the above separation is very approximate, and there is likely to be contamination in both samples. Taking into account that the separation is uncertain, especially for sources with no MIPS 24\microns fluxes, we estimate that approximately 30 to 50\percent in the red source catalog are likely to be AGB stars, and 50 to 70\percent are likely to be YSOs. Figure \ref{fig:separated} shows the angular distribution of all the AGB and YSO candidates, and shows that the two populations are reasonably well separated by the simple criteria provided in this section.

\subsection{Angular Distribution of AGB stars}

\label{sec:expfit}

\begin{figure}[p]
\begin{center}
\includegraphics[width=5.0in]{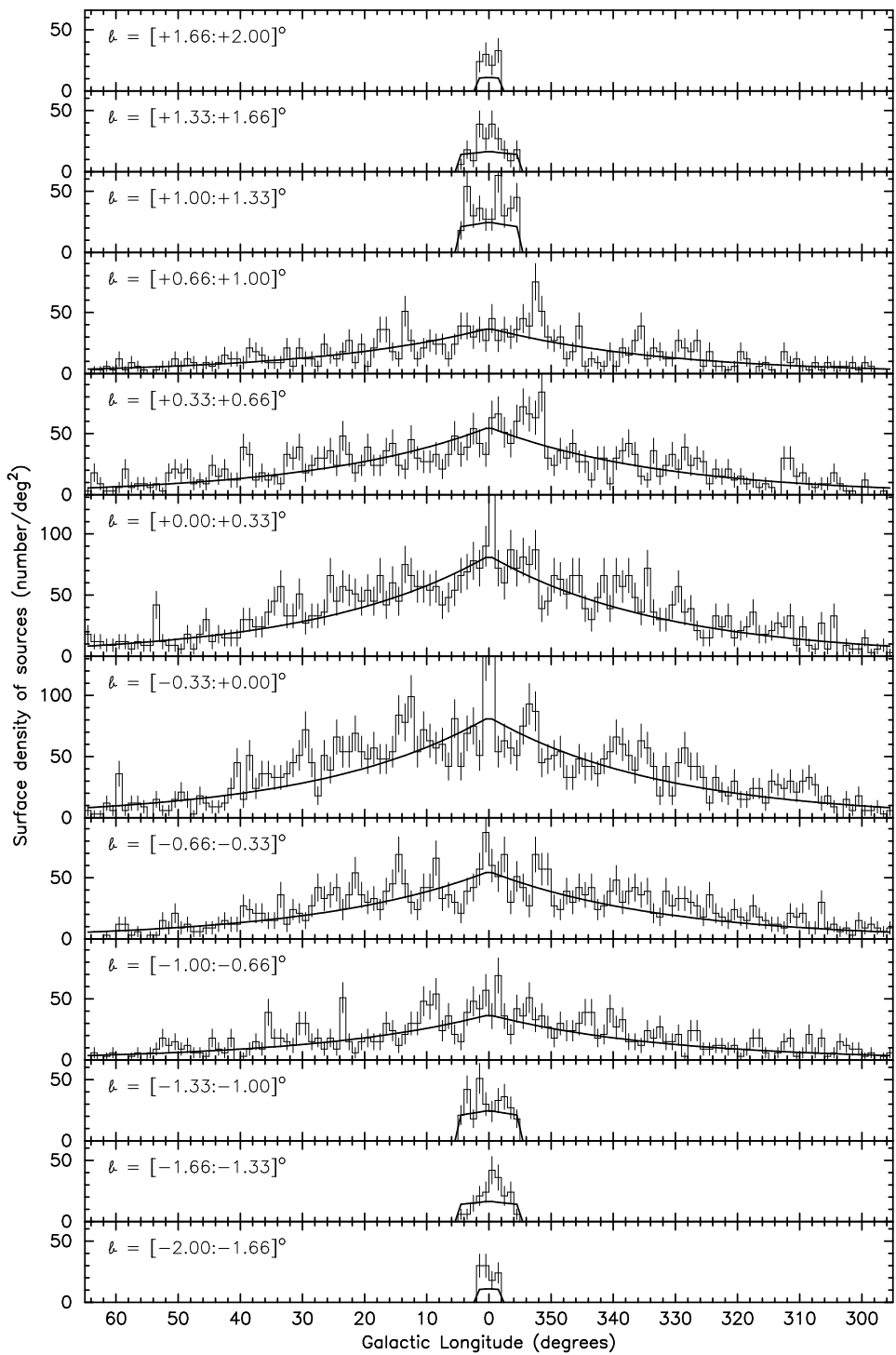}\\
\caption{Surface density of the candidate AGB stars as a function of Galactic longitude and latitude (histogram) with the best fit analytical distribution shown (solid line). The error bars are Poisson uncertainties. The analytical form of the best fit as well as the best fit parameters are given in \S\ref{sec:expfit}.\label{fig:agb_dist}}
\end{center}
\end{figure}

While the angular distribution of candidate YSOs appears highly clustered, the distribution of candidate AGB stars appears to be fairly smooth. In fact, the longitude and latitude distribution can be well approximated by a simple function of the form:
\begin{equation}
\Sigma(\ell,b) = \Sigma_0\,\exp{\left(-|\ell|/\ell_0\right)}\,\exp{\left(-|b|/b_0\right)}
\end{equation}

To fit this function to the distribution of sources, the surface density of all candidate AGB stars was estimated in 130 longitude bins ($1$\degrees wide) and 9 latitude bins ($0.33$\degrees high), and Poisson uncertainties were calculated for each bin. The distribution of AGB stars is shown in Figure~\ref{fig:agb_dist}, with the best fit overplotted. The best fit parameters were found to be $\Sigma_0 = 100\pm3$\,deg$^{-2}$, $\ell_0=14.1\pm0.3$\degrees, and $b_0=0.418\pm0.014$\degrees, and the reduced $\chi^2$ of the fit was found to be 1.479. This fit is therefore a good statistical description of the surface density of the AGB stars inside the GLIMPSE area. However, we note that the fit tends to underestimate the surface density of AGB stars for $|b|>1.33$\degrees. In addition, the central  peak in the density of AGB stars seen in Figure \ref{fig:separated} cannot be explained by such a simple function. Finally, we emphasize that this function is only an approximation to the surface density of AGB stars that are present in the red source catalog presented in this paper, rather the surface density of all AGB stars in the GLIMPSE Catalogs, which is likely to be much higher due to the large number of \sagb stars that are likely to have $[4.5]-[8.0]<1$.

\subsection{Distance-Luminosity sensitivity}

\label{sec:distances}

\begin{figure}[p]
\begin{center}
\includegraphics[width=3in]{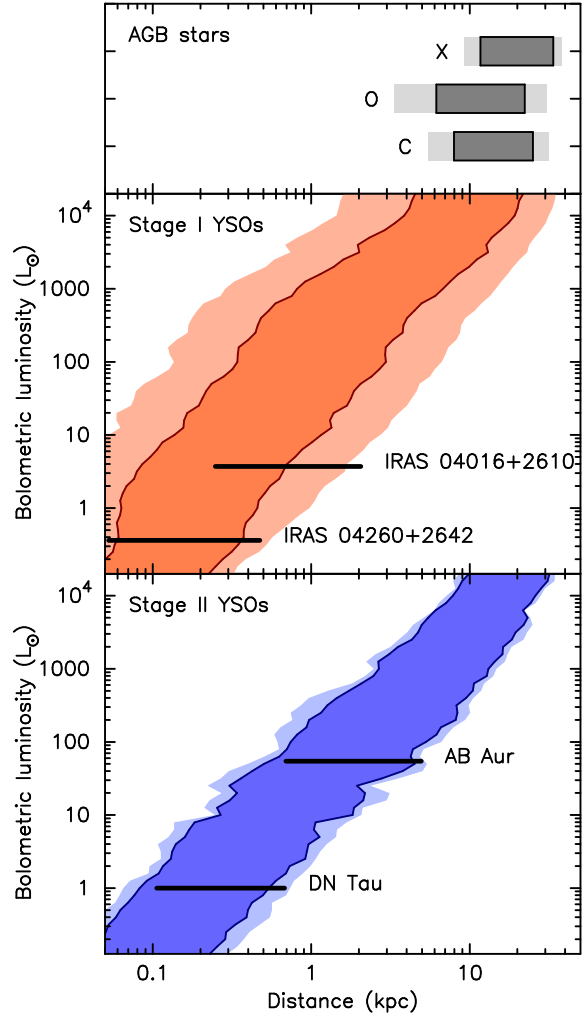}\\
\caption{\textit{Top} - The region surrounded by the solid line is the distance range in which an AGB star with average [4.5] and [8.0] magnitudes would need to be in order to be present in our red source catalog. The average magnitudes were derived for LMC AGB stars with $[4.5]-[8.0]\ge1$. X indicates \xagb stars, C indicates C-rich \sagb stars, and O indicates O-rich \sagb stars. The lighter region outside this corresponds to the 1-$\sigma$ standard deviation in the average magnitudes. \textit{Bottom} - The luminosity limit as a function of distance for embedded YSOs (Stage I), and for non-embedded YSOs with protoplanetary disks (Stage II). The solid lines enclose the distance ranges corresponding to the average [4.5] and [8.0] magnitudes, and the lighter region outside this corresponds to the 1-$\sigma$ standard deviation in the average magnitudes. The distance ranges in which a few known objects would be detectable are also shown.\label{fig:lum_dist}}
\end{center}
\end{figure}

In order to determine the distance range in which the different types of AGB stars in our red source catalog might lie, the average [4.5] and [8.0] magnitudes of \xagb and \sagb stars in the LMC with $[4.5]-[8.0]\ge1$ were computed, as well as their standard deviation. The distance range where these `average' AGB stars would satisfy the brightness criteria for inclusion in the red source catalog (Equation [\ref{eq:brightness}]) was then determined, assuming an approximate extinction-distance relation of 1.9\,mag/kpc \citep{Allen:73} and the visual to mid-infrared extinction conversion described in \S\ref{sec:agb}. The resulting distance ranges are shown in dark gray in Figure \ref{fig:lum_dist}, while the light gray ranges show the variation in the distance ranges corresponding to the standard deviations on the average magnitudes.

For YSOs, the models from \citet{Robitaille:06:256} were used. Average [4.5] and [8.0] magnitudes and standard deviations on these were determined as a function of bolometric luminosity for two evolutionary stages: the embedded phase and the protoplanetary disk phase, or typical Stage I and II models using the `Stage' definition from \citet{Robitaille:06:256}. The typical Stage I models were taken to be those with $\dot{M}_{\rm env}/M_\star=5\times10^{-5}\rightarrow2\times10^{-4}$\peryr, and the typical Stage II models were taken to be those with no infalling envelope and $M_{\rm disk}/M_\star=0.005\rightarrow0.02$. Only models with viewing angles between $30$\degrees and $60$\degrees were used, and the fluxes were taken to be those inside a $\sim15,000$\au aperture. The resulting distance ranges are shown encompassed by the solid lines in Figure \ref{fig:lum_dist}.

For the Stage I models, the standard deviations in the [4.5] and [8.0] magnitudes were typically less than $2.5$\mag (or a factor of $10$ in flux). These standard deviations are large because mid-infrared wavelengths are very sensitive to geometrical effects, such as viewing angle \citep{Whitney:03:1049}. For Stage II models, the standard deviations are less than $1$\mag. In Figure \ref{fig:lum_dist}, the scatter in the distance range corresponding to the standard deviations on the average [4.5] and [8.0] magnitudes are shown in lighter color outside the solid lines.

This analysis shows that the faintest AGB stars should be detectable no closer than 3\kpc (closer AGB stars would exceed the saturation level), while the brightest AGB stars should be visible even at the far-side of the Galaxy. \xagb stars should be visible further away on average than \sagb stars. YSOs should also be visible throughout the Galaxy: for example, 1\lsun YSOs should be visible up to $\sim0.8-1$\kpc; 100\lsun Stage I YSOs should be visible from $200$\pc to nearly $10$\kpc, while Stage II YSOs of the same luminosity (such as AB Aur) should be visible from $0.5$ to
$5$\kpc; and $10^4$\lsun YSOs should be visible at the far side of the Galaxy.

\subsection{Spectral Energy Distributions}

\begin{figure}[t]
\begin{center}
\includegraphics[width=6.0in]{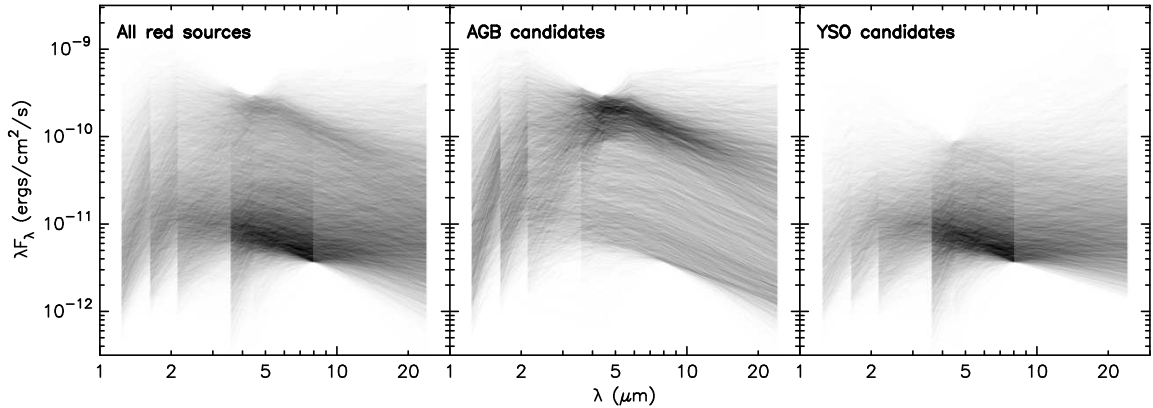}\\
\caption{\textit{Left} - stacked SEDs for all sources in the red source catalog. \textit{Center} - stacked SEDs for all AGB candidates. \textit{Right} - stacked SEDs for all YSO candidates.\label{fig:seds_all}}
\end{center}
\end{figure}

\begin{figure}[p]
\begin{center}
\includegraphics[width=6.0in]{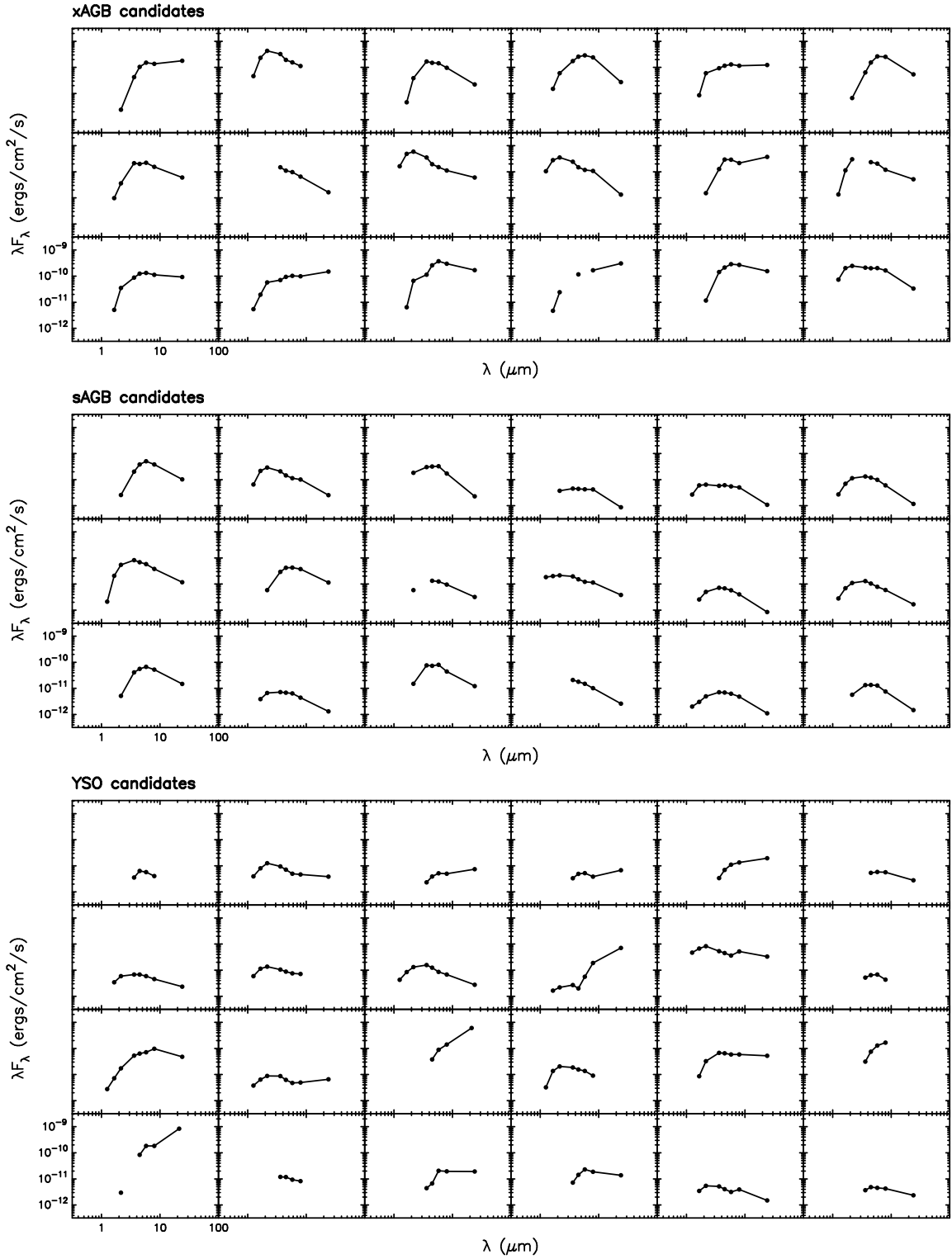}\\
\caption{Example SEDs of sources selected randomly from the \xagb, \sagb, and YSO candidates.\label{fig:seds_examples}}
\end{center}
\end{figure}

Figure \ref{fig:seds_all} shows the stacked SEDs of all sources in the red source catalog. A higher density of sources is present both at the bright and the faint end, consistent with the color-magnitude diagrams in Figure \ref{fig:color_red}. Also shown are the stacked SEDs of only the AGB star candidates, and only the YSO candidates. Figure \ref{fig:seds_examples} shows typical SEDs for sources in the \xagb, \sagb, and YSO categories. The YSO SEDs show a much larger variation in SED shape than the AGB stars.

\section{Summary}

\label{sec:summary}

We have compiled a flux-limited catalog of nearly 19,000 sources in the Galactic plane that are intrinsically red at mid-infrared wavelengths, using data from the \textit{Spitzer} GLIMPSE~I and II surveys and IRAC observations of the Galactic center. The sources were required to satisfy the brightness and quality selection criteria from equation (\ref{eq:brightness}) and (\ref{eq:quality}) to improve the reliability of the red source catalog (\S\ref{sec:initial_selection}), and were required to have $[4.5]-[8.0]\ge1$ to be considered `red' (\S\ref{sec:red_selection}). The latter criterion was determined to be the most straightforward and reproducible way of selecting red sources (\S\ref{sec:qualitative_selection}). In particular, the interstellar extinction law is mostly flat between 4.5 and 8.0\microns, meaning that any red color between the two bands is likely to be intrinsic rather than due to interstellar extinction.

Independent mosaic photometry was performed on all sources selected by these criteria to validate the GLIMPSE 4.5 and 8.0\microns photometry, and each source was visually examined to ensure that the independent flux could be trusted. Sources for which reliable independent photometry could not be performed were rejected from the red source catalog. In addition, sources for which the independent flux was more reliable than the original GLIMPSE flux and for which the new fluxes no longer satisfied the brightness and color selection criteria from equations (\ref{eq:brightness}) and (\ref{eq:color}) were rejected from the red source catalog. In total, 18,949 red sources satisfied all the selection criteria and were determined to have reliable photometry. The final catalog is given in Table~\ref{tab:red}, and includes JH\ks, IRAC, and MIPS 24\microns magnitudes. It is at least 65\percent complete (for point sources) and close to 100\percent reliable, meaning that every source in the catalog is a genuine red source.

The near- and mid-infrared color-magnitude and color-color distribution of the red sources was presented (\S\ref{sec:spatial}). One particular feature of the IRAC color-magnitude distribution of the red sources was the presence of two distinct populations, one peaking at bright [4.5] and [8.0] magnitudes ($[8.0]<6.5$ and $[4.5]<8.0$), and one peaking at faint [4.5] and [8.0] magnitudes ($[8.0]>8$ and $[4.5]>8.5$). The bright peak consists mostly of bright AGB stars, while the remaining sources consist mostly of fainter AGB stars and YSOs.
Using simple color and magnitude selection criteria, the red sources were separated into three distinct populations (\S\ref{sec:decontamination}): `extreme' or `obscured' C- and O-rich AGB stars (\xagb stars), `standard' C- and O- rich AGB stars (\sagb stars), and YSOs.

The angular distribution, near- and mid-infrared colors, and magnitudes of these three populations were found to differ. The \xagb stars appear to be seen at least as far as the Galactic center, show a rapid drop-off with Galactic longitude, and as expected for distant sources, their near-infrared colors therefore suggest higher values of interstellar extinction than the other populations. The \sagb stars show a somewhat shallower dropoff with Galactic longitude, and show less reddening than the \xagb stars, consistent with their closer distance. While they are also expected to be seen at least as far as the Galactic center (\S\ref{sec:distances}), no concentration of sources is seen at the Galactic center itself, but this could be an artifact due to the requirement for a valid MIPS 24\microns flux for the classification (\S\ref{sec:decontamination}). Finally, the YSOs show a shallow dropoff with Galactic longitude, and their distribution is highly clustered, unlike the two populations of AGB stars. The approximate separation of the three populations suggests that approximately $30-50$\percent of sources in the red source catalog are likely to be AGB stars, and approximately $50-70$\percent are likely to be YSOs. The fraction of red sources that are galaxies and PNe was found to be very small, on the order of a few \percent at most (\S\ref{sec:pne} and \S\ref{sec:galaxies}).

The AGB stars in the red source catalog are likely to form one of the largest samples of mid-infrared selected AGB stars in the Galaxy to date, with over 7,000 AGB star candidates. In particular, the coverage of the GLIMPSE~II region at two epochs has allowed us to uncover over a thousand sources with significant variability ($>0.3$\mag) at 4.5 and/or 8.0\microns, which we identify as \xagb stars with Mira variability. These represent one fifth of all red sources in the GLIMPSE~II region. Of all the AGB stars in the Galaxy that fall in the GLIMPSE survey area (but are not necessarily detected by \textit{Spitzer}), the red source catalog is likely to only contain a small fraction ($\sim1$\percent) of all sAGB stars, but may contain up to a quarter of all xAGB stars.

In parallel, over 11,000 YSO candidates have been uncovered. These do not provide a \textit{complete} picture of Galactic star formation as seen by \textit{Spitzer}. For example, the red source catalog does not include blended sources, extended sources, point source YSOs for which excess emission at 4.5\microns due to H$_2$ and CO bandhead 
emission from outflows makes the $[4.5]-[8.0]$ too blue for the selection criterion used in this paper, and sources that are so embedded that they are not detected at IRAC wavelengths. Nevertheless, it is a first step towards a study of star formation - as seen by \textit{Spitzer} - on a Galactic scale, and is to date the largest consistently selected sample of YSOs in the Milky-Way. These thousands of YSOs trace many previously known and some previously unknown sites of star formation, including large star formation complexes, smaller star formation regions, and dark clouds. From this perspective, the red source catalog can be thought of not only as a large sample of AGB stars and YSOs, but as the most detailed map to date of the birth sites of intermediate and massive stars in the Galactic plane.

\acknowledgements

We wish to thank the referee, Neal Evans, for helpful suggestions which helped improve this paper, Sundar Srinivasan for providing source lists of LMC AGB stars, John Grave for useful discussions regarding variability in the GLIMPSE~II region, and Martin Groenewegen for advice regarding LMC and Milky-Way AGB stars.
This work was supported by a Scottish Universities Physics Alliance Studentship (TPR), NASA grants to the {\it Spitzer} Legacy GLIMPSE, GLIMPSE~II, and GLIMPSE~3D projects (EC, MRM, BLB, BAW, RI, MC, RAB), the \emph{Spitzer} Theoretical Research Program (BAW, TPR), the {\it Spitzer} Fellowship Program (RI), JPL/\textit{Spitzer} grants 1275467 and 1276990 (RI), the NASA Theory Program (BAW; NNG05GH35G), and a PPARC studentship (KGJ).
This research is based on observations made with the {\it Spitzer} Space Telescope, which is operated by the Jet Propulsion Laboratory, California Institute of Technology under a contract with NASA; has made use of data products from the Two Micron All Sky Survey (2MASS), which is a joint project of the University of Massachusetts and the Infrared Processing and Analysis Center/California Institute of Technology, funded by NASA and the National Science Foundation; and has used the VizieR database of astronomical catalogues \citep{Ochsenbein:00:23}.

%%%%%%%%%%%%%%%%% References %%%%%%%%%%%%%%%%%

\bibliography{}
 
%%%%%%%%%%%%%%%%% Figures %%%%%%%%%%%%%%%%%

\clearpage

\begin{deluxetable}{crrrrrrrrrrrrrrrrrrc}
\rotate
\tabletypesize{\tiny}
\tablewidth{0pt}
\tablecaption{Final red source catalog\label{tab:red}}
\tablehead{
& \multicolumn{4}{c}{Celestial coordinates\tablenotemark{a}} & & \multicolumn{3}{c}{2MASS} & & \multicolumn{4}{c}{GLIMPSE Catalog} & & \multicolumn{5}{c}{This paper} \\
\cline{2-5} \cline{7-9} \cline{11-14} \cline{16-20}
\colhead{Source name} & \colhead{$\ell$} & \colhead{$b$} & \colhead{$\alpha$ (J2000)} & \colhead{$\delta$ (J2000)} & & \colhead{J} & \colhead{H} & \colhead{\ks} & & \colhead{[3.6]} & \colhead{[4.5]} & \colhead{[5.8]} & \colhead{[8.0]} & & \colhead{[4.5]} & \colhead{[8.0]} &  \colhead{MSX E} & \colhead{[24.0]} & \colhead{Flag\tablenotemark{b}}
}
\startdata
SSTGLMC G000.0000$+$00.1611  &    0.0000 &    0.1611 &  266.2480 &  $-$28.8521 &  &  \nodata &  \nodata &  \nodata &  &    9.78 &    9.19 &    8.64 &    8.05 &  &    9.13 &    8.17 &  \nodata &  \nodata & AA  \\
SSTGLMC G000.0000$-$00.4342  &    0.0000 &   $-$0.4342 &  266.8295 &  $-$29.1617 &  &  \nodata &  \nodata &   11.73 &  &    8.29 &    6.67 &    5.37 &    4.30 &  &    6.89 &    4.31 &  \nodata &    0.89 & IA  \\
SSTGLMC G000.0031$-$00.5072  &    0.0031 &   $-$0.5072 &  266.9029 &  $-$29.1969 &  &  \nodata &  \nodata &  \nodata &  &   12.62 &   10.09 &    7.60 &    5.68 &  &   10.37 &    5.69 &  \nodata &    1.41 & IA  \\
SSTGLMC G000.0046$+$01.1431  &    0.0046 &    1.1432 &  265.2992 &  $-$28.3321 &  &   10.64 &    9.87 &    9.40 &  &    7.63 &    6.67 &    5.78 &    5.04 &  &    6.64 &    5.03 &  \nodata &    4.50 & AA  \\
SSTGLMC G000.0058$+$00.1527  &    0.0058 &    0.1528 &  266.2596 &  $-$28.8516 &  &   13.62 &   11.98 &   10.73 &  &    9.10 &    8.53 &    7.85 &    6.78 &  &    8.56 &    6.76 &  \nodata &    3.59 & AA  \\
SSTGLMC G000.0083$-$00.4818  &    0.0084 &   $-$0.4817 &  266.8810 &  $-$29.1792 &  &  \nodata &  \nodata &  \nodata &  &    9.74 &    8.79 &    7.85 &    7.19 &  &    8.88 &    7.18 &  \nodata &    3.97 & AA  \\
SSTGLMC G000.0085$+$00.1542  &    0.0085 &    0.1543 &  266.2597 &  $-$28.8484 &  &  \nodata &  \nodata &   12.57 &  &    8.31 &    6.86 &    5.78 &    5.31 &  &    6.87 &    5.29 &  \nodata &    2.94 & AA  \\
SSTGLMC G000.0098$+$00.1625  &    0.0098 &    0.1626 &  266.2525 &  $-$28.8430 &  &  \nodata &   13.39 &  \nodata &  &    8.34 &    7.59 &    7.07 &    6.32 &  &    7.50 &    6.26 &  \nodata &    2.54 & AA  \\
SSTGLMC G000.0106$-$00.7315  &    0.0106 &   $-$0.7314 &  267.1274 &  $-$29.3063 &  &   11.31 &    8.97 &    7.77 &  &  \nodata &    7.18 &  \nodata &    6.07 &  &    7.21 &    6.01 &  \nodata &    2.60 & AA  \\
SSTGLMC G000.0110$-$01.0237  &    0.0110 &   $-$1.0236 &  267.4151 &  $-$29.4564 &  &  \nodata &  \nodata &  \nodata &  &    9.25 &    8.32 &    7.27 &    6.30 &  &    8.44 &    6.24 &  \nodata &    2.15 & AA  \\

\nodata & \nodata & \nodata & \nodata & \nodata & & \nodata & \nodata & \nodata & & \nodata & \nodata & \nodata & \nodata & & \nodata & \nodata & \nodata & \nodata \\
\enddata
\tablenotetext{a}{These coordinates are set to the average position of the source at 4.5 and at 8.0\microns. This position may differ slightly from the `official' GLIMPSE position in cases where PSF fitting was used to determine the flux of the source if the position of the source was adjusted to obtain a better residual.}
\tablenotetext{b}{This column lists two characters, which are flags for 4.5 and 8.0\microns respectively. A = GLIMPSE Catalog magnitudes are in agreement with the independent magnitudes calculated in this paper. I = the independent magnitudes calculated in this paper should be trusted over the GLIMPSE Catalog magnitudes.}
\tablecomments{The full table is available on request (\texttt{tr9@st-andrews.ac.uk}), and will be available as an electronic table as part of the AJ publication. The zero-magnitude fluxes assumed throughout this paper are: $F_\nu(J)=1594$\Jy, $F_\nu(H)=1024$\Jy, $F_\nu($K$_{\rm s})=666.7$\Jy, $F_\nu(3.6\mu$m$)=280.9$\Jy, $F_\nu(4.5\mu$m$)=179.7$\Jy, $F_\nu(5.8\mu$m$)=115.0$\Jy, $F_\nu(8.0\mu$m$)=64.13$\Jy, $F_\nu($MSX~E$)=8.75$\Jy, $F_\nu(24.0\mu$m$)=7.14$\Jy.}
\end{deluxetable}

\clearpage

\begin{deluxetable}{crrrrrrrcrrcc}
\rotate
\tabletypesize{\footnotesize}
\tablewidth{0pt}
\tablecaption{Red sources from the GLIMPSE~II region with photometry at two epochs\label{tab:variables}}
\tablehead{ &\multicolumn{4}{c}{Celestial coordinates\tablenotemark{a}} & & \multicolumn{2}{c}{Epoch 1} & & \multicolumn{2}{c}{Epoch 2} & & \\
\cline{2-5} \cline{7-8} \cline{10-11}
\colhead{GLIMPSE source name} & \colhead{$\ell$} & \colhead{$b$} & \colhead{$\alpha$ (J2000)} & \colhead{$\delta$ (J2000)} & & \colhead{[4.5]} & \colhead{[8.0]} & & \colhead{[4.5]} & \colhead{[8.0]} & &  \colhead{Variable\tablenotemark{b}}}
\startdata
SSTGLMC G000.0046$+$01.1431  &    0.0046 &    1.1432 &  265.2992 &  $-$28.3321 &  &    6.67 &    5.04 &  &    6.89 &    5.22 & & N \\
SSTGLMC G000.0106$-$00.7315  &    0.0106 &   $-$0.7314 &  267.1274 &  $-$29.3063 &  &    7.18 &    6.07 &  &  \nodata &    5.17 & & Y \\
SSTGLMC G000.0110$-$01.0237  &    0.0110 &   $-$1.0236 &  267.4151 &  $-$29.4564 &  &    8.32 &    6.30 &  &    8.34 &    6.32 & & N \\
SSTGLMC G000.0115$-$01.2781  &    0.0115 &   $-$1.2780 &  267.6664 &  $-$29.5864 &  &    8.25 &    6.84 &  &    8.30 &    6.87 & & N \\
SSTGLMC G000.0127$-$01.2146  &    0.0127 &   $-$1.2146 &  267.6045 &  $-$29.5529 &  &    6.66 &    5.53 &  &    6.58 &    5.51 & & N \\
SSTGLMC G000.0177$-$01.8424  &    0.0177 &   $-$1.8423 &  268.2291 &  $-$29.8687 &  &   10.13 &    8.04 &  &   10.10 &    8.04 & & N \\
SSTGLMC G000.0371$+$01.6473  &    0.0371 &    1.6475 &  264.8339 &  $-$28.0370 &  &    6.79 &    5.20 &  &  \nodata &    4.70 & & Y \\
SSTGLMC G000.0408$-$00.7197  &    0.0408 &   $-$0.7196 &  267.1336 &  $-$29.2744 &  &   10.80 &    6.00 &  &   10.77 &    5.99 & & N \\
SSTGLMC G000.0540$-$00.7328  &    0.0540 &   $-$0.7327 &  267.1543 &  $-$29.2698 &  &    7.93 &    6.91 &  &    8.09 &    6.90 & & N \\
SSTGLMC G000.0566$-$00.7363  &    0.0567 &   $-$0.7363 &  267.1593 &  $-$29.2693 &  &    9.42 &    7.90 &  &    9.40 &    7.83 & & N \\

\nodata & \nodata & \nodata & \nodata & \nodata & & \nodata & \nodata & & \nodata & \nodata & & \nodata \\
\enddata
\tablenotetext{a}{These coordinates are set to the average position of the source at 4.5 and at 8.0\microns, as in Table~\ref{tab:red}.}
\tablenotetext{b}{Whether the magnitudes for the two epochs differ by at least 0.3\mag at either (or both) 4.5\microns or 8.0\microns.}
\tablecomments{The full table is available on request (\texttt{tr9@st-andrews.ac.uk}), and will be available as an electronic table as part of the AJ publication.}
\end{deluxetable}

\end{document}